\documentclass[a4paper,11pt]{article}
\pdfoutput=1
\usepackage{jcappub}
\usepackage{caption}
\usepackage{hyperref}
\usepackage{amsmath}
\usepackage{mathtools}
\usepackage{upgreek}
\usepackage{tablefootnote}
\usepackage{soul}
\usepackage{color}
\usepackage{verbatim}
\usepackage{float}
\usepackage{bm}
\usepackage[usenames,dvipsnames,svgnames,table]{xcolor}
\usepackage{graphicx}
\usepackage{amssymb}
\usepackage{epstopdf}
\graphicspath{ {./Figures/} }
\usepackage{dcolumn}
\usepackage{bm}
\usepackage[caption=false]{subfig}
\usepackage{mhchem}
\usepackage[normalem]{ulem}
\usepackage{multirow}
\usepackage{booktabs}
\usepackage{tikz}
\usetikzlibrary{positioning}

\newcommand{\be}{\begin{eqnarray}}

\newcommand{\ee}{\end{eqnarray}}

\usepackage[export]{adjustbox}% http://ctan.org/pkg/adjustbox

\newcommand{\bsk}{\boldsymbol{k}}
\newcommand{\bl}{\boldsymbol{\ell}}

\newcommand{\kmin}{k_\mathrm{min}}
\newcommand{\kmax}{k_\mathrm{max}}
\newcommand{\lmin}{\ell_\mathrm{min}}
\newcommand{\lmax}{\ell_\mathrm{max}}
\newcommand{\ld}{\ell_\mathrm{D}}
\newcommand{\local}{\mathrm{local}}
\newcommand{\equil}{\mathrm{equil}}
\newcommand{\ortho}{\mathrm{ortho}}

\begin{document}

% \preprint{APS/123-Q ED}

\title{Fundamental limits on constraining primordial non-Gaussianity }

\newcommand{\VSI}{Van Swinderen Institute for Particle Physics and Gravity,\\ University of Groningen,
Nijenborgh 4, 9747 AG Groningen, The Netherlands}
\newcommand{\LION}{Lorentz Institute for Theoretical Physics, Leiden University, Leiden University, 2333 CA, The Netherlands }
\newcommand{\UVA}{Institute of Physics, University of Amsterdam, Amsterdam, 1098 XH, The Netherlands}
\newcommand{\kavli}{Institute of Astronomy and Kavli Institute for Cosmology, Madingley Road, Cambridge, UK, CB3 0HA}

\newcommand{\IPMU}{Kavli Institute for the Physics and Mathematics of the Universe (WPI), The University of Tokyo Institutes for Advanced Study (UTIAS), The University of Tokyo, Kashiwa, Chiba 277-8583, Japan}

\author[a]{Alba Kalaja,}
\emailAdd{a.kalaja@rug.nl}

\author[a]{P.\ Daniel Meerburg,}

\author[b,c]{Guilherme L. Pimentel,}

\author[d,e]{William R. Coulton}

\affiliation[a]{\VSI}
\affiliation[b]{\LION}
\affiliation[c]{\UVA}
\affiliation[d]{\kavli}
\affiliation[e]{\IPMU}
\date{\today}

\abstract{We study the cosmic variance limit on constraining primordial non-Gaussianity for a variety of theory-motivated shapes. We consider general arguments for 2D and 3D surveys, with a particular emphasis on the CMB. A scale-invariant $N$-point correlator can be measured with a signal-to-noise that naively scales with the square root of the number of observed modes. This intuition generally fails for two reasons. First, the signal-to-noise scaling is reduced due to the blurring of the last scattering surface at short distances. This blurring is caused by the combination of projection and damping, but the loss of signal is not due to exponential decay, as both signal and noise are equally damped. 
Second, the behavior of the $N$-point correlator in the squeezed and collapsed (for $N>3$) limits can enhance the scaling of the signal-to-noise with the resolution, even with a reduced range of momenta probing these limits. 
We provide analytic estimates for all $N$-point correlators. We show that blurring affects equilateral-like shapes much more than squeezed ones. We discuss under what conditions the optimistic scalings in the collapsed limit can be exploited. 
Lastly, we confirm our analytical estimates with numerical calculations of the signal-to-noise for  local, orthogonal and equilateral bispectra, and local trispectra. We also show that adding polarization to intensity data enhances the scaling for equilateral-like spectra.
}

\maketitle
\pagebreak
\section{Introduction}
\begin{figure}[t!]
    \centering
    \includegraphics[width=0.5\textwidth]{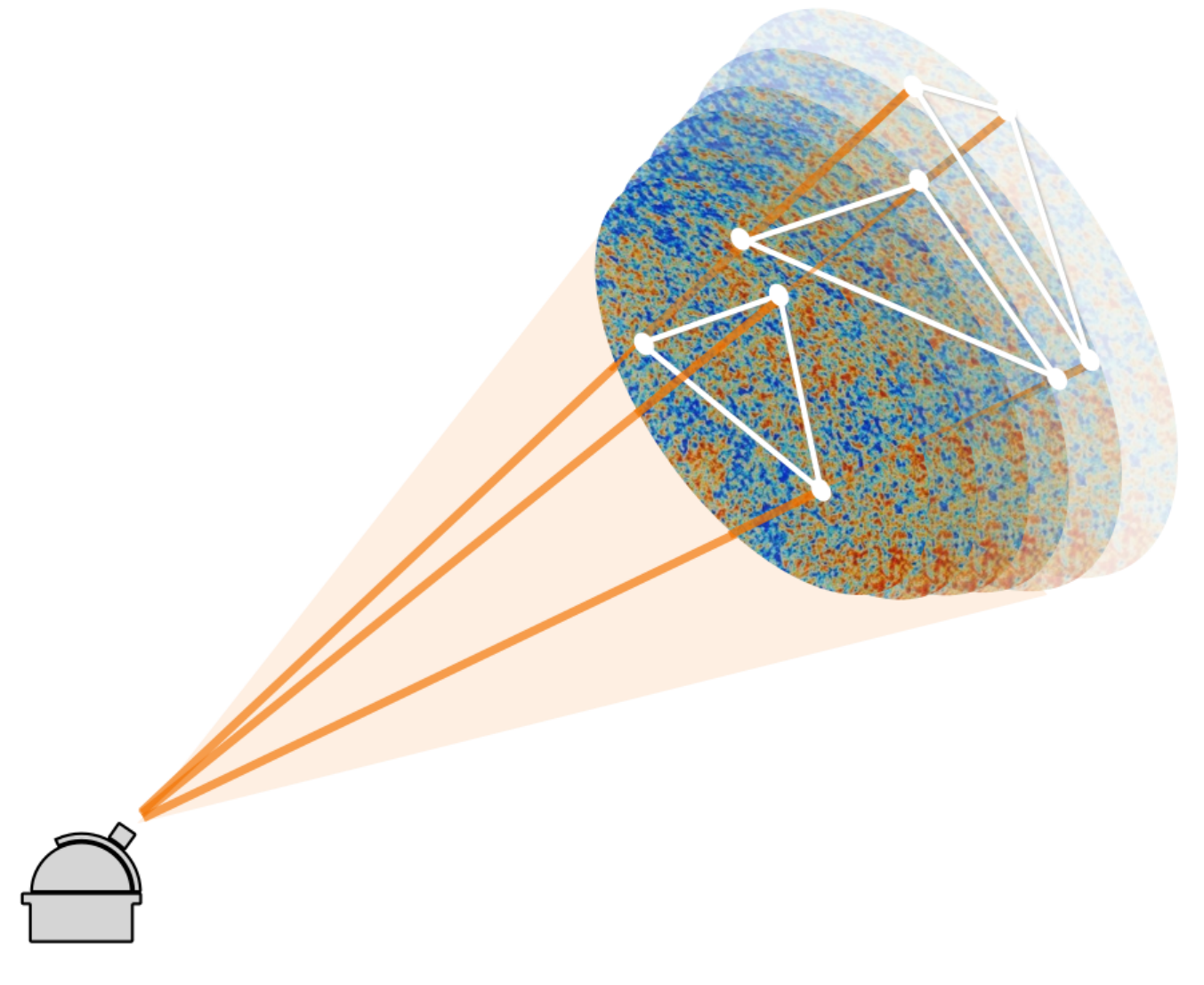}
    \caption{The blurriness of the last scattering surface at short scales washes out our ability to detect primordial non-Gaussianity. In the figure, we show a cartoon that largely exaggerates this effect for illustration purposes. When we measure the correlations within a triangle in the sky, we must average over all possible triangles along the same line of sight. This averaging reduces the signal to noise ratio as a function of the survey resolution $\ell_{\rm max}$.}
    \label{fig:damping_thickness}
\end{figure}

The statistics of primordial fluctuations imprinted in the Cosmic Microwave Background (CMB) and visible in the late-time clustering of galaxies is our most important window into the very early universe. 
A measurement and characterization of the primordial fluctuations is necessary to improve our understanding of a putative period of inflation, shedding light on its microscopic origin. As future experiments achieve higher sensitivity and resolution, increasing the capacity to detect statistics beyond the two-point correlation function, i.e. non-Gaussianity \cite{meerburg:png_whitepaper}, both CMB~\cite{ade:so_forecast,matsumura:litebird_forecast,abazajian:cmb4_forecast,hanany:pico} and large scale structure (LSS) missions~\cite{amendola:euclid,marshall:lsst,aghamousa:desi,green:wfirst,bacon:ska,dore:spherex} will face challenges associated to removing foregrounds and taking into account non-primordial effects (see e.g.~\cite{coulton:minimizing_gravlens_bispectrum,Hill_2018} and \cite{Darwish:2020prn}). Besides these systematic challenges, we run into limitations determined by the nature of the tracer field. For example, the CMB is fundamentally limited by two factors: 
\begin{itemize}
    \item the last scattering surface is two-dimensional, thus observed anisotropies are a 2D projection of the 3D fluctuations;
    \item the thickness of the last scattering surface requires that we average over all projections along the line of sight, effectively blurring the non-Gaussianity on small scales, as shown in Fig.~\ref{fig:damping_thickness}.
\end{itemize}
In this paper we build a qualitative and quantitative understanding of how these limitations affect the signal-to-noise. We tackle the problem in two ways: first, using a scale invariant spectrum of non-Gaussianity as theoretical prior, we derive the scaling relation of the signal-to-noise ratios ($S/N$) with the number of measured fluctuations (modes), i.e. the inverse of the resolution of a given experiment. Secondly, we perform a numerical Fisher analysis to test the consistency of our theoretical predictions. In both cases, we will assume cosmic variance limited observations and neglect non-primordial sources that could hinder a detection of primordial non-Gaussianity. 

\begin{figure}[h!]
    \centering
    \includegraphics[width=0.7\textwidth]{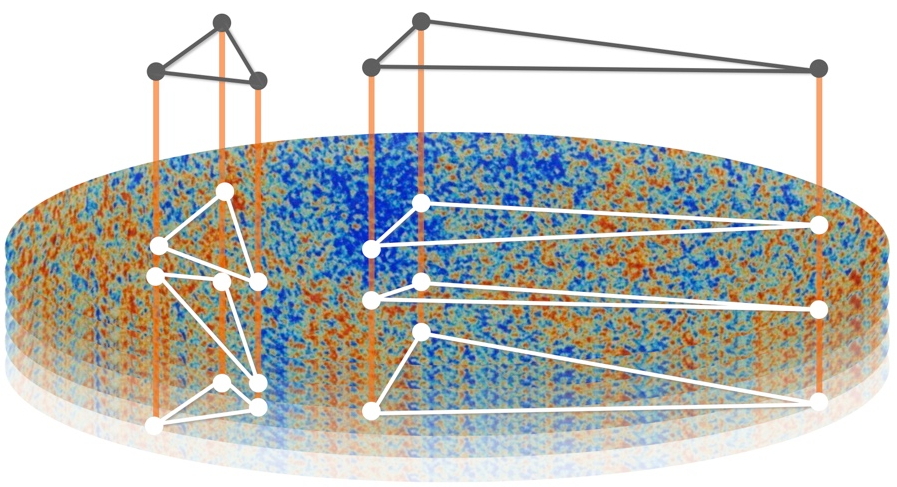}
    \caption{For shapes of non-Gaussianity peaked around the squeezed limit, blurriness does not reduce the $S/N$ ratio much. This is because within the same line of sight the possible triangles do not change shape drastically. For equilateral-like non-Gaussianity, we have to average over triangles with various shapes and the blurriness reduces the $S/N$ significantly.}
    \label{fig:equil_local_damping}
\end{figure}

In order to contextualize our motivation, let us consider the bispectrum of temperature anisotropies. First, neglecting the effect of diffusion damping on small scales, a simple estimate shows that the $S/N$ scales linearly with the maximal measured multipole $\lmax$. This follows from scale invariance of the bispectrum. However, when the damping effect is taken into account in the transfer function, the $S/N$ shows a much poorer scaling and grows with the square root of $\lmax$. This result was derived in Ref.~\cite{bartolo:ng_from_recombination} in the context of equilateral non-Gaussianity. Here, we show that, at least in these simple estimates, the problem gets worse at higher point correlation functions, to the extreme that the $S/N$ converges (for five- and higher point correlation functions), even if all the CMB modes are measured. We argue that this result is a general consequence of the blurring of non-Gaussianity at small scales. 
Furthermore, in view of recent advances in the theory of cosmological correlators, for our theoretical estimates we go beyond equilateral and local bispectra analysis, and consider more general shapes.
In particular, around squeezed or collapsed limits---when there is a hierarchy between distance scales being correlated---, non-Gaussianity exhibits features similar to those of a particle detector in a collider experiment, potentially probing new particles that decay into the primordial fluctuations~\cite{arkanihamed:cosmo_collider, lee:ng_particle_detector}. 
While the exact scaling is determined by the full shape of a given primordial $N$-point correlation function---or, equivalently, the bispectrum, trispectrum, etc.\footnote{A slightly confusing terminology is that $N$-point correlation functions are referred to as $(N-1)$-spectra. Thus the $3$-point correlation function is the bispectrum, the $4$-point correlation function is the trispectrum, etc.}---
we find that the leading scaling can be estimated by computing only the squeezed and collapsed limits of correlators. This greatly simplifies calculations of $S/N$ ratios, allowing us to adapt the effect of the blurriness and to extend the argument to all ($N-1$)-spectra. Despite the reduced phase space in restricted kinematics, for certain shapes, the presence of damping at small scales turns out to have little influence on the $S/N$ scaling.
Intuitively, as shown in Fig.~\ref{fig:equil_local_damping}, when large scale perturbations are correlated with small ones, the signal is well defined since the possible triangles keep the same shape along the line of sight. On the other hand, a signal coming from the correlation of perturbations with a similar scale (which is below the damping scale) receives contributions from different shapes due to the thickness of the last scattering surface. Thus the average over all possible triangles along the line of sight produces a blurriness that reduces the $S/N$. 
We show that, indeed, equilateral-like shapes have a more reduced $S/N$ than shapes with different wavelengths modes correlated. In particular, we find that squeezed bispectra and collapsed trispectra have a large $S/N$ scaling, in the case where it is sourced by the exchange of very light particles during inflation. These results were already shown in  Ref.~\cite{babich:bispectrum_cmb_polarization} and Refs.~\cite{kogo:angular_trispectrum, bordin:higher_spin_cmb_statistics} for the bispectrum and the trispectrum respectively. However, the enhanced scaling of the trispectrum needs to be interpreted properly, within a specific model generating the shapes~\cite{creminelli:estimators_local_ng}. We discuss this later in the paper. 

We will confirm our heuristic estimates for the bispectrum and the trispectrum with a full-sky Fisher estimate. Here, we will use simplified templates of non-Gaussianity, that still capture the relevant physics of the {\it ab initio} shapes. We also include $E$-mode polarization in our estimates, and show that they increase the $S/N$ ratio, sometimes parametrically in $\lmax$. Our interpretation of this finding is that polarization knows about the velocity field around recombination, and effectively increases the dimensionality of the CMB to be slightly above 2D. A similar effect is found in Ref.~\cite{coulton:rayleigh_scattering_info}, where Rayleigh scattering is included as a tracer of primordial fluctuations. Adding data to the primary CMB temperature modes only improves the $\lmax$ scaling of shapes for which the scaling is not already optimal (e.g. squeezed shapes, which already reach mode-counting scaling). 

The paper is organized as follows: in Sec.~\ref{sec:ng_from_inflation}, we briefly review the shapes of the correlators in the squeezed and collapsed limits; in Sec.~\ref{sec:theory_snr}, we show the theoretical estimates of the $S/N$; in Sec.~\ref{sec:numerical_analysis}, we outline the numerical Fisher analysis and compare the results with the theoretical ones; in Sec.~\ref{sec:discussion_conclusion}, we discuss the results of the paper and future prospects. In the appendices we include additional technical details and derivations: App.~\ref{app:cmb_fullsky},~\ref{app:cmb_flatsky} and~\ref{app:snr_derivation} contain a brief review of CMB statistics and $S/N$ derivation; in App.~\ref{app:damping_cmb} we explicitly show the effect of diffusion damping, and in App.~\ref{app:multiple_squeezed_collapsed} we derive multiple squeezed and collapsed limits.

\vskip 10 pt
\noindent \textbf{Results} The signal-to-noise ratio is a function of various quantities: $\ell_{\rm max}$ ($\ell_{\rm min}$), the smallest (largest) angular resolutions of the survey; $f_{\rm NL}$, the size of the non-Gaussianity\footnote{In this case, $f_{\rm NL}$ is a general parameter, not to be confused with the amplitude of the bispectrum.}; $f_{\rm sky}$, the available fraction of the sky being measured; and the details of the shape--local, equilateral, degree of correlation function, etc. For weak non-Gaussianity, the dependence on $f_{\rm NL}$ and $f_\mathrm{sky}$ is very simple. They appear as overall factors in the signal-to-noise. The dependence on $\ell_{\rm max}$, $\ell_{\rm min}$ (or $k_{\rm max}$, $k_{\rm min}$ for a $3$D survey) is more complicated. A detailed analysis of the various shapes gives us
\begin{equation}
    \left(\frac{S}{N}\right)^2\sim (f_\mathrm{sky} f_{\rm NL})^2 \ell_{\rm max}^p ~,
\end{equation}
where we omit numerical factors and the $\ell_{\rm min}$ dependence\footnote{In principle, $\lmin$ is limited by the fraction of the sky observed. For simplicity, we assume $\lmin$ to have a fixed value.}. The power $p$ depends on the shape of the non-Gaussianity and whether $\ell_{\rm max}$ is above or below the damping scale $\ld$. 
We parametrize a large family of non-Gaussian shapes by their behavior around squeezed and collapsed limits, and find the corresponding scaling $p$ above and below the damping scale. The resulting values of $p$ are summarized in Tab.~\ref{tab:all_scalings}.
\vskip 10 pt
\noindent \textbf{Notation and conventions} The modulus of a vector is given by $|\bsk|\equiv k$ and $|\boldsymbol{\ell}| \equiv \ell$. We use $\bsk_{ij\dots n} =  \bsk_I+\bsk_j +\dots+\bsk_n$ and $\boldsymbol{\ell}_{ij\dots n} =  \boldsymbol{\ell}_I+\boldsymbol{\ell}_j +\dots+\boldsymbol{\ell}_n$. 

\noindent We denote the primordial curvature perturbation $\zeta (\bsk) = \zeta_{\bsk}$. The various moments of the fluctuations are defined as
\begin{equation}
\label{eq:moments_curvature}
\begin{split}
    &\langle \zeta_{\bsk_1} \zeta_{\bsk_2}\rangle = (2\pi)^3 \delta^{(3)}(\bsk_{12})P_\zeta (k_1),\\
    &\langle \zeta_{\bsk_1} \zeta_{\bsk_2}\zeta_{\bsk_3}\rangle = (2\pi)^3 \delta^{(3)}(\bsk_{123})B_\zeta (k_1,k_2,k_3),\\
    &\langle \zeta_{\bsk_1} \zeta_{\bsk_2}\zeta_{\bsk_3}\zeta_{\bsk_4}\rangle = (2\pi)^3 \delta^{(3)}(\bsk_{1234})T_\zeta (\bsk_1,\bsk_2,\bsk_3,\bsk_4),\\
    &\langle \zeta_{\bsk_1} \zeta_{\bsk_2}\cdots\zeta_{\bsk_N}\rangle = (2\pi)^3 \delta^{(3)}(\bsk_{12\dots N}) F_\zeta (\bsk_1,\bsk_2,\dots, \bsk_N).
\end{split}
\end{equation}

\noindent In our numerical computations we consider a flat $\Lambda$CDM cosmology, with cosmological parameters in accordance with the latest \textit{Planck} results~\cite{planck2018:cosmological_parameters}, summarized in Tab.~\ref{tab:cosmology}. 
\begin{table}[h!]
    \centering
    \begin{tabular}{c  c  c}
     \toprule
     & $\boldsymbol{\Lambda}$\textbf{CDM} \textbf{parameters} &\\
     \midrule
        H$_0 = 67.66$ &  $\Omega_\mathrm{b} h^2 = 0.02242$ & $\sum$m$_\nu = 0.06$\\
        $\Omega_\mathrm{k} = 0$ &  $\Omega_\mathrm{c} h^2 = 0.11993$  & $\tau = 0.0561$\\
         $n_s = 0.9665$& $A_\mathrm{s} = 2.1056\times 10^{-9}$  & $r = 0$\\
        \bottomrule
    \end{tabular}
    \caption{Best-fit Planck parameters (specifically, Tab.~2 of Ref.~\cite{planck2018:cosmological_parameters} with $TT$, $TE$, $EE+$low$E+$lensing+BAO) used in our numerical computations.}
    \label{tab:cosmology}
\end{table}

\section{Non-Gaussianity from inflation}
\label{sec:ng_from_inflation}
Cosmic inflation provides a compelling mechanism for the generation of primordial fluctuations -- as the universe expands and flattens, quantum mechanical vacuum fluctuations are stretched to cosmological distances. During the past decades and more recently from the \textit{Planck} mission~\cite{planck2018:cosmological_parameters}, measurements of the CMB fluctuations have shown consistency with the predictions of slow-roll models of inflation: superhorizon primordial fluctuations are nearly scale-invariant, adiabatic and nearly Gaussian. 
While the single field slow-roll model predicts a small amount of non-Gaussianity~\cite{maldacena:sfsr_ng,acquaviva:sfsr_ng}, coming from gravitationally mediated self-interactions of the inflaton, other models of inflation predict stronger signals. 
Different physical processes, mediated by particles with masses of order the Hubble scale during inflation, give rise to distinctive signatures in the non-Gaussian signal. The precision calculation of such signatures is a rich subject, much akin to the area of scattering amplitudes in particle physics. In certain kinematical configurations, such as squeezed and collapsed limits, the shape of non-Gaussianity reveals information about the mass and the spin of particles mediating the interactions among the curvature fluctuations~\cite{chen:equil_trispectrum,baumann:supersymmetry_early_universe,arkanihamed:cosmo_collider,lee:ng_particle_detector,meerburg:png_whitepaper}. Generally, the $N$-point correlation functions of the primordial curvature fluctuations are given by
\begin{equation}
  \langle\zeta_{\bsk_1}\zeta_{\bsk_2}\cdots \zeta_{\bsk_N}\rangle=\delta^{(3)}(\bsk_{12\dots N}) F_\zeta(\bsk_1,\dots,\bsk_N)
\end{equation}
Assuming rotation and translation invariance, the degrees of freedom of the scalar function $F_\zeta(\bsk_1,\cdots,\bsk_N)$ will be reduced to $3(N-2)$.
All the information about the amplitude of non-Gaussianity and its specific shape is encoded in $F_\zeta$. Below, we focus on the shape of $F_\zeta$ and we leave out the coefficients in front of it, which are related to the amplitude. 

Initially we ignore the angular dependence of the correlators and focus on investigating how the volume of phase space affects the $S/N$\footnote{The angular dependence of correlators as a function of momenta most clearly gives the signature of a spinning particle mediating the interaction among curvature fluctuations.}. We then examine specific kinematical regimes, i.e. angular limits, and show how these limits determine the dominant scaling of the $S/N$. 

When all momenta are of similar size, then we correlate wavelengths generated at the same time during inflation, and we expect to probe contact, self-interactions of the inflaton. Around this kinematics, non-Gaussianities are devoid of features, and have a scaling consistent with the symmetries of the inflationary background\footnote{ The scaling of $F_\zeta$ has a factor of $k^{3(N-1)}$ for an $N$-point function. The extra factor of $k^{-3}$ comes from the momentum conserving delta function, rendering the correlator to be scale invariant.},
\begin{equation}
   F_{\zeta}(k_1\sim k_2\sim\cdots \sim k_N)\sim \frac{1}{k_1^{3(N-1)}}\, .
\end{equation}
The details of the shape become clearer as we dial the various momenta, generating a hierarchy between the times in which the various fluctuations are sourced during inflation. 

There are two interesting kinematical limits of the correlator that generate a hierarchy between momenta:
\begin{itemize}
    \item {\bf Squeezed limits---}When one of the momenta is much smaller than all the others, we probe correlations between fluctuations generated at two different times during inflation. If there are new particles $\sigma$, perhaps massive, that couple to the inflaton, then their imprints become clearest in the squeezed limit, acting as mediators between fluctuations generated at earlier and later times (see left diagram in Fig.~\ref{fig:n_point}).  In the squeezed limit, a single leg probes the later time, and due to momentum conservation, the long mediator has the same wavelength as said external fluctuation. A convenient way of parametrizing the general behavior of the squeezed limit is\footnote{Throughout, we will drop anything but the degrees of freedom relevant to $F_\zeta$. }
\begin{equation}
   \lim_{k_1\ll k_{2,\dots,N}} F_\zeta(k_1,k_2, \dots, k_N)\sim \frac{1}{k_1^3 k_2^{3(N-2)}} \left(\frac{k_1}{k_2}\right)^{\Delta}+\dots,
    \label{eq:n_point_squeezed}
\end{equation}

where $\dots$ are subleading terms, suppressed by powers of the small ratio $k_1/k_2$. The leading order exponent $\Delta$ is related to the mass of the particle in Hubble units. In detail, when inflation is very well approximated by de Sitter space (zero-order in slow-roll), then $\Delta=3/2-\sqrt{9/4-m_\sigma^2/H^2}$. We will be agnostic about the specific microscopic origin of $\Delta$, but it is useful to keep in mind that it is related to the mass of a mediator particle.

\begin{figure}[t]
\centering
\includegraphics[width=0.7\textwidth]{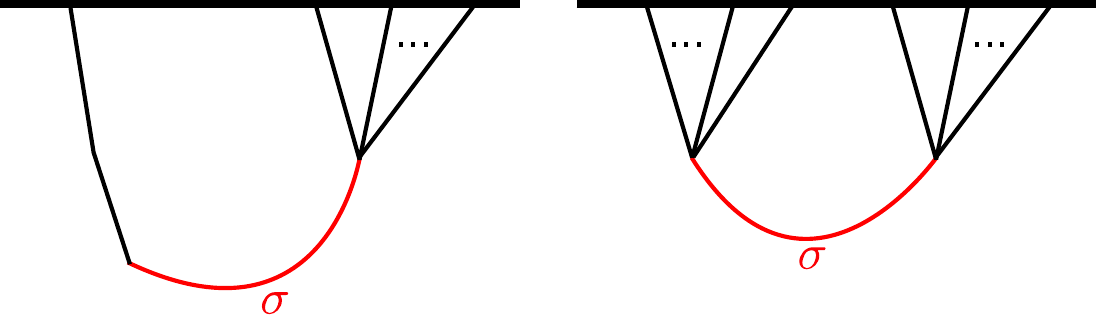}
\caption{\textit{Left}: Squeezed $N-$point functions. Cartoon of physical process sourcing squeezed non-Gaussianity. Time grows along the vertical direction and the horizontal line marks the end of inflation. The squeezed momentum corresponds to a fluctuation with very long wavelength, sourced at early times. A particle $\sigma$ mediates a correlation between the long mode and the $N-1$ short modes sourced at later times. The scaling behavior of the correlator is related to the mass (in Hubble units) of $\sigma$. \textit{Right}: A collapsed $N-$point function configuration. We assume that all momenta on each ``bundle" have similar magnitude, but the ``left" and ``right" momenta can be different. In the figure, left and right momenta are sourced at similar times, but the correlation is mediated by $\sigma$, which is generated earlier. In the limit where the exchanged momentum is much smaller than the momenta on each side of the diagram, the correlation function will exhibit scaling controlled by the mass of the exchanged particle.}
\label{fig:n_point}
\end{figure}
    \item {\bf Collapsed limits---}Instead of squeezing a side of a polygon, we can correlate two collections of modes generated late in inflation mediated by a pair produced at early times, with long wavelength. In order to probe such process, we consider ``collapsing" a polygon (at least a quadrilateral, i.e. the trispectrum) by making a diagonal very small\footnote{Note that permutation symmetries of $N$-point functions imply correlators must be the same for two polygons with very different shapes, obtained by permuting the order of the sides. As a consequence of this, it is reductive to assume the collapsed limits arise only when a diagonal becomes small. In turn, collapsed configurations correspond also to cases where none of the geometrical diagonals vanish and it is better to say that a subset of momenta adds up to zero. The results we obtain in Sec.~\ref{sec:theory_snr} therefore apply more broadly, but are derived in the limit where the configuration is chosen such that the diagonal vanishes.}. More generally, we can bundle the momenta in two groups, as shown in the right panel in  Fig.~\ref{fig:n_point}. Then, denoting the ``left" momenta $\{k_1,\cdots k_M\}\sim k_L$ and the ``right" momenta $\{k_{M+1},\cdots k_N\}\sim k_R$ (with $M>1$),  the mass of the mediator particle will be read off by the scaling $\Delta$ in the collapsed limit $k_I\ll k_L, k_R$
\begin{equation}
   \lim_{k_I \ll k_L, k_R} F_\zeta(k_I, k_L, k_R)\propto \frac{1}{k^{3(M-1)}_L k^3_I k^{3(N-M-1)}_R} \left(\frac{k_I^2}{k_R k_L}\right)^{\Delta}+\dots,
    \label{eq:n_point_collapsed}
\end{equation}
where $k_I\equiv k_1+\cdots +k_M$. Notice that both the squeezed limit and the collapsed limit give access to similar processes, but the squeezed limit requires linear mixing of $\sigma$ with the inflaton, while more general interactions are accessible via collapsed limits. Moreover, momentum conservation still gives certain phase space in the various configurations of left and right momenta, while keeping the diagonal collapsed. 

\end{itemize} 

Finally, for large enough values of $\Delta$, the non-Gaussianity quickly decays as any hierarchy between momenta is generated. These shapes generally arise from contact self-interactions of the primordial fluctuations during inflation, and are referred to as {\it equilateral} non-Gaussianities\footnote{In the effective field theory of single-field inflation, there are two distinct self-interactions of the inflaton that produce distinct patterns of non-Gaussianity in the bispectrum. One of them is referred to as the equilateral shape, while the other is referred to as the orthogonal shape. Nonetheless, both of those shapes probe the physics of self-interactions of the curvature.}. 

\begin{figure}[t!]
\centering
\includegraphics[width=0.6\textwidth]{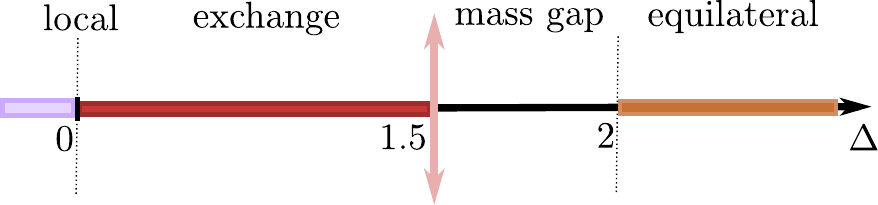}
\caption{The value of $\Delta$ encodes interesting new physics of particles generating long range forces during inflation. The plot above shows the relation between the mass of a new particle and the scaling of the squeezed limit of the bispectrum. For higher masses, the values of $\Delta$ become complex (see the pink arrows), and the non-Gaussian signal develops an oscillating pattern. Notice the ``mass gap" for $3/2<\Delta<2$ -- the simplest inflationary models predict that the leading contribution to the bispectrum should not lie on that range.}
\label{fig:delta_mass_range}
\end{figure}

A summary of the discussion above is presented in Fig.~\ref{fig:delta_mass_range}. A measurement of $\Delta$ provides spectroscopic information about what sets up the initial conditions, correlating  primordial fluctuations along different scales. 
In the simplest inflationary scenarios, $0<\Delta<3/2$ signals the presence of a light (in Hubble units) mediator particle setting long range correlations among the inflaton particles. Values $\Delta\ge 2$ probe self-interactions of the inflaton and include equilateral non-Gaussianity. In many models, as for example in slow-roll, quasi-single field inflation, the value of $\Delta$ in the squeezed bispectrum is often the same as that in the collapsed trispectrum. A notable exception to this rule is graviton exchange. In the collapsed limit, the masslessness of the graviton is evident, and the scaling of the trispectrum is $\Delta=0$. In the squeezed bispectrum, we only exchange the static Newtonian potential between the fluctuations, which is effectively a contact interaction, and the scaling is that of an equilateral shape, with $\Delta=2$. 

Our discussion of shapes of non-Gaussianity has not been exhaustive, but it will suffice for the analysis below. We will assume general values of $\Delta$, while not necessarily the same in the squeezed vs. collapsed limit. We should mention that other shapes leave very interesting features, related to spins of the particles, bursts of particle production, different inflationary mechanisms, alternative scenarios for sourcing the initial conditions etc. We leave the extension of our analysis to other shapes for future work.

\section{Theoretical signal-to-noise ratio estimation}
\label{sec:theory_snr}
In this section, we give analytical estimations of the $S/N$ of non-Gaussianity from general $N$-point correlation functions. Our estimates assume that our only limitation in directly probing non-Gaussianity is the finite number of modes in the Universe, i.e. cosmic variance.

In principle, we expect the $S/N$ of the amplitude of non-Gaussianities to scale with the square root of the total number of modes in a survey. For a 3D survey, as in the case of an LSS mission, the total number of modes is proportional to $N_{3\mathrm{D}}\sim 1/r_{\rm min}^3$, with $r_{\rm min}$ the smallest observed scale within the survey, or equivalently in Fourier space $N_{3\mathrm{D}}\sim\kmax^3$. 

For the CMB, the survey is confined to a 2D surface, therefore we expect the total number of modes to be proportional to $N_{2\mathrm{D}}\sim\lmax^2$, with  $\lmax$ the largest accessible multipole. However, some effects that are intrinsic to our probes, such as the finite thickness of the last scattering surface for the CMB, drastically change the amount of information that we can access, underestimating it with respect to mode counting.

\begin{figure}[t!]
    \centering
    \includegraphics[width=0.7\textwidth]{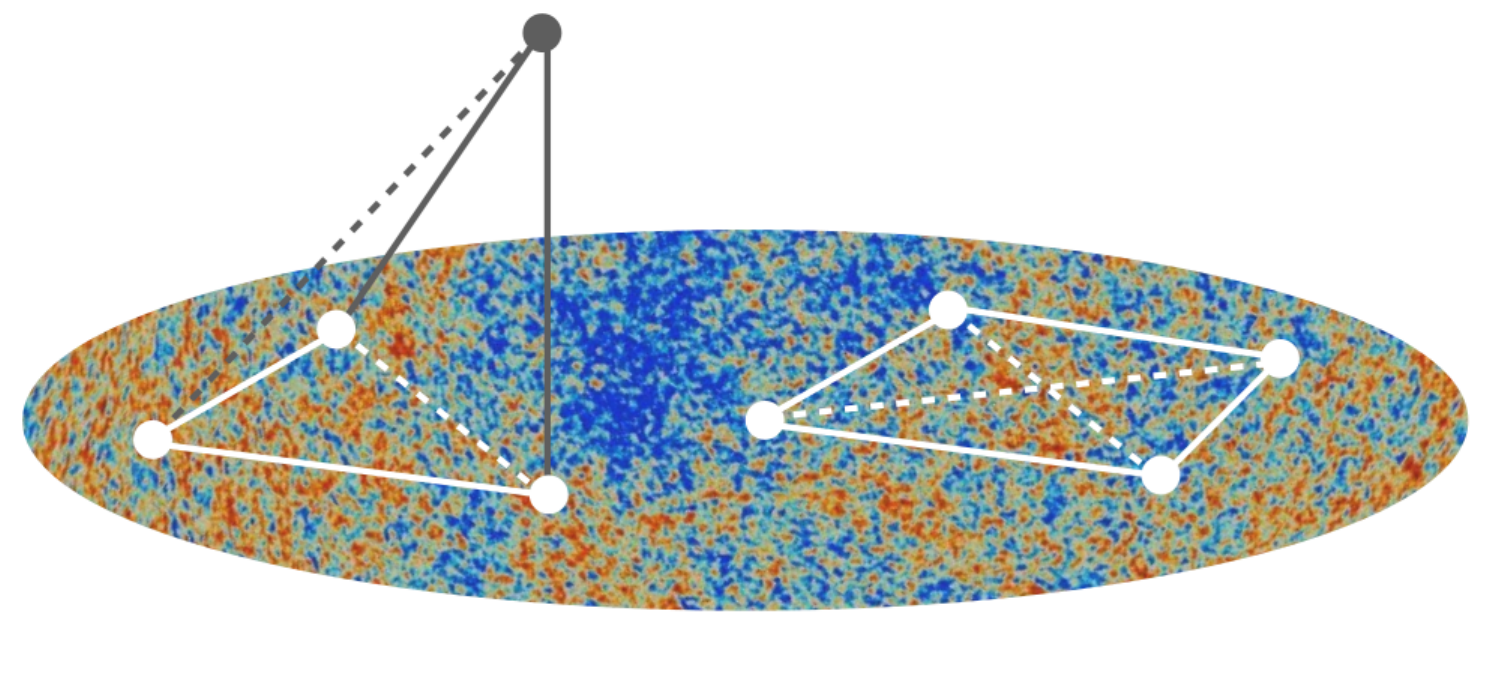}
    \caption{The last scattering surface is a 2D surface where fluctuations are projected.}
    \label{fig:damping2d3d}
\end{figure}

In Sec.~\ref{subsec:general_snr_estimates}, we give a general overview of the scaling of $S/N$ in terms of $\kmax$ ($\lmax$) for a 3D (2D) survey. Then, we show how damping affects the $S/N$ scaling for a CMB survey for general $(N-1)$-spectra. Here, it is interesting to point out an important distinction between the bispectrum and the trispectrum. In order to specify a triangle in 2D vs. 3D, three pieces of data (the sides of the triangle) are needed. In that sense, there is no loss of freedom in going from a tomogram to a projection. This is not true for a quadrilateral, which would require six pieces of data (four sides, two diagonals) in 3D, while requiring five pieces of data (four sides, one diagonal) in 2D\footnote{Strictly speaking, given four sides and one diagonal, there are two possible quadrilaterals in two dimensions, one concave and one convex. However, it remains true that there is no freedom to dial the size of the second diagonal arbitrarily.} (see e.g. Fig.~\ref{fig:damping2d3d}). This restriction does not seem to play an important role in the asymptotic estimates below, but it might be important in analyzing features or more detailed signatures of new physics in the trispectrum, where having the freedom to change the various shapes of the quadrilateral becomes important.

Finally, in Sec.~\ref{subsec:squeezed_limit} and~\ref{subsec:collapsed_limit} we study the scaling properties of the $S/N$ in the squeezed and collapsed limits respectively, highlighting the dependence on the specific shape considered. 

For the sake of clarity, we use the flat-sky approximation (see e.g. App.~\ref{app:cmb_flatsky}) to describe the CMB statistics and we limit our analysis to temperature perturbations. These estimations are far from being rigorous, but they convey the message without lengthy calculations.

\subsection{General estimation: the damping effect on the \texorpdfstring{$S/N$}{S/N}}
\label{subsec:general_snr_estimates}
For a 3D survey, the $(S/N)^2$ of an $N$-point correlation functions is given by
\begin{equation}
      \begin{split}
       \left(\frac{S}{N}\right)^2_{(N)}\sim \prod_{i=1}^N\left(\int \mathrm{d}^3k_i\right)\,\delta^{(3)}(\bsk_{12\dots N})\frac{F^2_\zeta(k_1,k_2,\dots,k_N)}{P(k_1)P(k_2)\cdots P(k_N)}.
    \end{split}
    \label{eq:snr_lss_nspectrum_sim}
\end{equation}
Here we neglect all coefficients in front of the integral since we only focus on the scaling in terms of $\kmax$. 
In general, for a 3D survey the scale-invariant $(N-1)$-spectrum scales as $F\sim k^{-3(N-1)}$, therefore by replacing it in Eq.~\eqref{eq:snr_lss_nspectrum_sim}, we obtain
\begin{equation}
    \left( \frac{S}{N}\right)^2_{(N)} \sim \int^{\kmax} \mathrm{d}^{3(N-1)} k\, \frac{k^{-6(N-1)}}{k^{-3N}}\sim \int^{\kmax}\mathrm{d}k\, k^2  \sim \kmax^3,
    \label{eq:npoint_scaling_modecounting}
\end{equation}
where we used the Dirac-delta to integrate one of the variables out.

The same argument applies to a 2D survey, where the $(S/N)^2$ reads as
\begin{equation}
\begin{split}
    \left(\frac{S}{N}\right)^2_{(N)}\sim \prod_{i=1}^N\left(\int \mathrm{d}^2\ell_i\right)\,
    \delta^{(2)}(\bl_{12\dots N}) \frac{F^2(\ell_1,\ell_2,\dots,\ell_N)}{ C(\ell_1)C(\ell_2)\cdots C(\ell_N)},
\end{split}
\end{equation}
Here we distinguish between two regimes: one where diffusion damping is ineffective, namely where $\ell \ll \ld$ with $\ld \simeq 1300$ the damping scale, the other one where diffusion damping becomes dominant, i.e. $\ell \gg \ld$. In the former, the flat-sky angular $(N-1)$-spectrum scales as $F\sim \ell^{-2(N-1)}$ (for more details see App.~\ref{app:cmb_flatsky}). Consequently, the $(S/N)^2$ scales as
\begin{equation}
\label{eq:flatsky_npoint_snr}
    \left( \frac{S}{N}\right)^2_{(N)} \sim \int^{\lmax} \mathrm{d}^{2(N-1)} \ell \frac{\ell^{-4(N-1)}}{\ell^{-2N}}\sim \int^{\lmax} \mathrm{d}\ell \,\ell \sim \lmax^2.
\end{equation}
As one would intuitively expect, both 3D and 2D estimates are proportional to the number of modes per survey, and the results are general for any $(N-1)$-spectra. 

The situation changes drastically in the $\ell \gg \ld$ regime, namely when the effect of the radiation transfer function, which we previously neglected, is incorporated in the estimation. As we show in detail in App.~\ref{app:damping_cmb}, the effect of the transfer function can be included by replacing  $\ell^2\rightarrow \ell^2(\ell/\ld)$. The damping in the exponential factor in the flat-sky transfer function~\eqref{eq:damping_transfer_function} cancel out in the $S/N$ (it appears equally in the numerator and the denominator), however the line-of-sight damping, which is sensitive to the thickness of the last scattering surface, modifies the scaling behavior of the $(N-1)$-spectra, and ultimately of the $S/N$. Indeed, if we add this effect in Eq.~\eqref{eq:flatsky_npoint_snr}, we obtain
\begin{equation}
    \left(\frac{S}{N}\right)^2_{(N)}\sim \ld^{N-2}\int d^{2(N-1)}\ell \,\frac{\ell^{-6(N-1)}}{\ell^{-3N}} \sim \ell_{\rm max}^{4-N}.
    \label{eq:n_snr_damping}
\end{equation}
The $N$ dependence implies that there is a intrinsic limitation in probing the shape of a general scale-invariant $(N-1)$-spectrum. While for $N=4$ the scaling is logarithmic,
\begin{equation}
    \left(\frac{S}{N}\right)^2_{(4)}\sim \ld^2\int^{\lmax} \mathrm{d}^6 \ell \frac{\ell^{-18}}{\ell^{-12}} \sim \log \lmax,
    \label{eq:trispectrum_snr_damping}
\end{equation}
the $(S/N)^2$ actually converges for $N>4$, where we expect
\begin{equation}
    \left(\frac{S}{N}\right)^2_{(N>4)}\sim A + \frac{B}{\lmax^p} \to A, ~~{\rm as}~\lmax\to \infty.
\end{equation}
As a consequence, diffusion damping along the line of sight puts a fundamental stop gap at how accurate we can distinguish signal from noise with $N>4$ correlation functions.
In this sense, looking for kinematical corners, where the scaling can improve, becomes crucial. In the following, we study the squeezed and collapsed limits of correlation functions. We will refer to the tracers with a reduced $S/N$ as \textit{damped tracers} while keeping in mind that the loss of information is due to a combination of projection and damping.

\subsection{The squeezed limit}
\label{subsec:squeezed_limit}
For the estimates made above, we used only the scaling properties of the $(N-1)$-spectrum, without making any assumptions about potential divergences around specific configurations or specific kinematic regimes. In this subsection, we estimate the scaling of the $(S/N)^2$ in the squeezed regime, stressing its dependence on the mass of mediators $\Delta$, hence on the shape of non-Gaussianity. We show that for certain values of $\Delta$, the scaling estimates in Sec.~\ref{subsec:general_snr_estimates} turn out to be an underestimation of the squeezed scaling. This result is trivial in the case of 3D surveys, however for high resolution CMB maps, well beyond the damping scale $\ld$, the squeezed limit of certain shapes produces a mode-counting scaling. 

We first describe in detail how to estimate the $(S/N)^2$ for the 3D and 2D $N$-point correlation function, the latter in the $\ell\ll\ld$ region. We then investigate the CMB bispectrum and trispectrum $(S/N)^2$ scaling in the damping region, $\ell\gg\ld$. These results are then generalized to the CMB $(N-1)$-spectrum.

Given the marked hierarchy between the momenta $k_1\ll k$, with $k = k_{2}\sim k_3 \sim \dots\sim k_N$, we can integrate $k_1$ from a minimum $k$-mode, $\kmin$, up to $k_1= c \kmax$, with $c\ll 1$. Then the $(S/N)^2$ becomes
\begin{equation}
    \begin{split}
        \left(\frac{S}{N}\right)^2_{(N)}&\sim \int_{\kmin}^{c\kmax}\mathrm{d}^3k_1 \int_{c\kmax}^{\kmax} \mathrm{d}^{3(N-2)}k\,\frac{k_1^{-6+2\Delta}k^{-6(N-2)-2\Delta}}{k_1^{-3}k^{-3(N-1)}}\\
        &\sim \int_{\kmin}^{c\kmax}\mathrm{d}k_1k_1^{2\Delta-1} \int_{c\kmax}^{\kmax} \mathrm{d}k\,k^{2-2\Delta},
    \end{split}
\end{equation}
where we used the squeezed $(N-1)$-spectrum defined in Eq.~\eqref{eq:n_point_squeezed}. 
The solution of the integral depends on the value of $\Delta$, but not on the correlation function considered, meaning that, given that shape, the results is universal for any bispectrum, trispectrum, etc. 

It is straightforward to show that, for general $\Delta > 0$, the scaling is given by mode-counting $\sim \kmax^3$. On the other hand, if $\Delta = 0$, which corresponds to the local type of non-Gaussianity, the $(S/N)^2$ shows a more favourable scaling with the number of modes, proportional to $\sim \kmax^3 \log (\kmax/ \kmin)$. 
In the extreme case where the $N$-point correlation function diverges faster than the local shape in the squeezed limit, namely for configurations with $\Delta<0$\footnote{$\Delta<0$ corresponds to mediator particles with imaginary mass, see e.g.~\cite{linde:hybrid_inflation,renauxpetel:destabilization_inflation} for inflationary scenarios where tachyonic instabilities are present for a few e-folds. Furthermore, the super-squeezed scaling might also imply a non-local theory~\cite{pajer:boostless_bootstrap_bispectrum}.}, we expect an enhancement of $\kmax^3$ by a factor of $(\kmax/\kmin)^{-2\Delta}$ .

The argument applies straight away to the CMB, with a simple $k^3\rightarrow \ell^2$ adjustment. In the region below the damping scale, the squeezed angular $(N-1)$-spectrum is given by
\begin{equation}
   \lim_{\ell_1\ll \ell \ll\ld} F_\zeta(\ell_1,\ell_2, \dots, \ell_N)\propto \frac{1}{\ell_1^2 \ell^{2(N-2)}} \left(\frac{\ell_1}{\ell}\right)^{2\Delta/3},
    \label{eq:cmb_n_point_squeezed}
\end{equation}
with $\ell = \ell_{2}\sim \ell_3 \sim \dots\sim \ell_N$. Since the integration is equivalent to the one outlined for the 3D $S/N$, we report only the results. We find
\begin{equation}
\begin{split}
    \left(\frac{S}{N}\right)^2_{(N)}\sim\begin{cases} \lmax^2, &\Delta > 0\\  \lmax^2 \log (\lmax/ \lmin), &\Delta=0.\end{cases}
\end{split}
\end{equation}
As in the 3D case, for $\Delta<0$, we expect an enhancement of $\lmax^2$ by a factor of $(\lmax/\lmin)^{-4\Delta/3}$.
\begin{table}[h!]
\renewcommand{\arraystretch}{1.2} 
\centering
\begin{tabular}{ccc}
& \multicolumn{2}{c}{Squeezed limit }  \\ \cline{2-3}
\multicolumn{1}{c|}{}  & \multicolumn{1}{c|}{3D} & \multicolumn{1}{c|}{2D ($\ell\ll \ld$)} \\
\cline{2-3}
\noalign{\vskip\doublerulesep{}
         \vskip-\arrayrulewidth}
\hline
\multicolumn{1}{|c|}{$\Delta < 0$} & \multicolumn{1}{c|}{$\kmax^3(\kmax/\kmin)^{-2\Delta}$} & \multicolumn{1}{c|}{$\lmax^2(\lmax/\lmin)^{-4\Delta/3}$} \\ \hline
\multicolumn{1}{|c|}{$\Delta = 0$} & \multicolumn{1}{c|}{$\kmax^3\log(\kmax/\kmin)$} & \multicolumn{1}{c|}{$\lmax^2\log(\lmax/\lmin)$}  \\ \hline 
\multicolumn{1}{|c|}{$\Delta > 0$} & \multicolumn{1}{c|}{$\kmax^3$} & \multicolumn{1}{c|}{$\lmax^2$}\\
\hline
\end{tabular}
\caption{General scaling behavior in the squeezed limit for the $(S/N)^2$ in 3D and 2D for $N$-point correlation functions.}
\label{tab:squeezed_3d2d_undamped}
\end{table}
The logarithmic enhancement of the scaling of the squeezed local non-Gaussianity is a well-known result in the literature, however the effect of the squeezed regime can be appreciated more at large $\ell\gg \ld$.

As already seen in Sec.~\ref{subsec:general_snr_estimates}, as we go to higher multipole values, the effect of diffusion damping along the line of sight affects the $S/N$ scaling (Eq.~\eqref{eq:n_snr_damping}). This result has immediate consequences for the statistics of the perturbations, limiting our capacity of extracting information on non-Gaussianity from CMB maps. 

Given the great interest in such observables, we take the squeezed limit of the damped angular bispectrum and trispectrum, before generalizing to the $(N-1)$-spectrum. We show that the signal greatly benefits from the squeezed limit, especially in the case of the bispectrum.

Using Eq.~\eqref{eq:n_snr_damping} for $N=3$, we see that the general scaling estimation is given by 
\begin{equation}
    \left(\frac{S}{N}\right)^2_{(3)}\sim \ld\int^{\lmax} d^{4}\ell \,\frac{\ell^{-12}}{\ell^{-9}} \sim \lmax,
    \label{eq:bispectrum_snr_damping}
\end{equation}
thus lower than mode-counting. Let us now take the squeezed bispectrum with the addition of the transfer function\footnote{Here, we neglect the exponential factor, given that it cancels out in the $S/N$ computation.}
\begin{equation}
     \lim_{\ell_1\ll \ell_2\sim \ell_3} B(\ell_1,\ell_2,\ell_3)\propto \frac{\ld^2}{\ell_1^3 \ell_2^3}\times \left(\frac{\ell_1}{\ell_2}\right)^{\Delta}. 
\end{equation}
Then, the $(S/N)^2$ scaling becomes
\begin{equation}
\begin{split}
    \left( \frac{S}{N}\right)^2_{(3)}\sim \ld \int_{\ld}^{c \lmax}\mathrm{d} \ell_1\ell_1^{2\Delta-2} \int^{\lmax}_{c \lmax} \mathrm{d} \ell_2\ell_2^{1-2\Delta},
\end{split}
\end{equation}
where we set the lower limit of integration to the damping scale $\ld$. We notice that there exist an interval of values $0<\Delta<1/2$, where the scaling with the number of modes improves with respect to $\lmax$ as
\begin{equation}
    \left( \frac{S}{N}\right)^2_{(3)}\sim\lmax\left(\frac{\lmax}{\ld}\right)^{1-2\Delta}.
\end{equation}
In particular, in the case of local non-Gaussianity, we recover mode-counting scaling $\sim \lmax^2$~\cite{babich:bispectrum_cmb_polarization}. 
On the contrary, for $\Delta>1/2$, the squeezed region becomes subdominant with respect to the damping scale and the scaling reduces to Eq.~\eqref{eq:bispectrum_snr_damping}. 

These results can be generalized to higher $(N-1)$-spectrum. In Eq.~\eqref{eq:trispectrum_snr_damping}, we showed that for the trispectrum, the $(S/N)^2$ scaling becomes worse with respect to the bispectrum result, being proportional to $\sim \log{\lmax}$. Performing a similar integration as done for the bispectrum, we have that the squeezed limit of the trispectrum is still less sensitive to non-Gaussianity than the bispectrum, as shown in Tab.~\ref{tab:squeezed_limit}. 

As for the $(N-1)$-spectrum, denoting with $\ell = \ell_2,\dots,\ell_N$, we obtain
\begin{equation}
    \left(\frac{S}{N}\right)^2_{(N)}=\int d^{2(N-2)}\ell \, d^2\ell_1 \,\frac{\ell_1^{-6+2\Delta} \ell^{-6(N-2)-2\Delta}}{\ell_1^{-3}\ell^{-3(N-1)}}\sim \ld^{-1+2\Delta}\ell_{\rm max}^{5-N-2\Delta}\,, \quad \text{for } 0<\Delta<1/2.
\end{equation}

Our analysis has a potential caveat --- we assumed that the scaling of the $N$-point function in the squeezed limit is ``equilateral"-like for the remaining hard momenta. The resulting $S/N$ can be an underestimate, if there is large signal when taking multiple squeezed limits. An example of interest, which we analyze numerically later on, is of the local trispectrum of $g^\local_\mathrm{NL}$ form,
\begin{equation}
    T_{\zeta}^{g_\mathrm{NL}}(\bsk_1,\bsk_2,\bsk_3,\bsk_4) =\frac{54}{25} g^\local_\mathrm{NL}[P_\zeta (k_2) P_\zeta (k_3)P_\zeta (k_4)+P_\zeta (k_1) P_\zeta (k_2)P_\zeta (k_4)+\text{2 perms.}]
\end{equation}
In that case, if we take a squeezed limit $k_1\ll k_{2,3,4}$ then the resulting function has permutations that look identical to the local bispectrum. In particular, we can consider the double squeezed limit $k_1\ll k_2 \ll k_{3,4}$, in which case our general estimates add up and we obtain 
\begin{equation}
     \left(\frac{S}{N}\right)^2_{g_{\rm NL}}\sim \int d^2\ell_{1} \ell_1^{-3}\int d^2\ell_{2} \ell_2^{-3} \int d^2\ell_{3} \sim \ell_{\rm max}^2 \, .
\end{equation}
In App.~\ref{app:multiple_squeezed_collapsed}, we generalize the idea of multiple squeezed limits to the damped $(N-1)$-spectrum and show that the scaling never exceeds mode-counting.
We conclude that the squeezed bispectrum likely remains the best observable for non-Gaussianity, especially to test light particles in the early Universe.

\begin{table}[h!]
\renewcommand{\arraystretch}{1.2} 
\centering
\begin{tabular}{c c c c}
& \multicolumn{3}{c}{Squeezed limit $\ell\gg \ld$}  \\ \cline{2-4}
\multicolumn{1}{c|}{}  & \multicolumn{1}{c|}{Bispectrum} & \multicolumn{1}{c|}{Trispectrum} & \multicolumn{1}{c|}{$(N-1)$-spectrum} \\
\cline{2-4}
\noalign{\vskip\doublerulesep{}
         \vskip-\arrayrulewidth}
\hline
\multicolumn{1}{|c|}{$\Delta< 1/2$} & \multicolumn{1}{c|}{$\lmax(\lmax/\ld)^{1-2\Delta}$} & \multicolumn{1}{c|}{$\ld(\lmax/\ld)^{1-2\Delta}$} & \multicolumn{1}{c|}{$\ld^{N-2}\lmax^{4-N}(\lmax/\ld^{N-2})^{1-2\Delta}$} \\ \hline 
\multicolumn{1}{|c|}{$\Delta = 1/2$} & \multicolumn{1}{c|}{$\ld^{2}\lmax\log(\lmax/\ld)$} & \multicolumn{1}{c|}{$\ld^{2}\log(\lmax/\ld)$} & \multicolumn{1}{c|}{$\ld^{2}\lmax^{4-N}\log(\lmax/\ld)$} \\ \hline 
\multicolumn{1}{|c|}{$\Delta > 1/2$} & \multicolumn{1}{c|}{$\lmax$} & \multicolumn{1}{c|}{$\ld^2(\lmax/\ld)^{1-2\Delta}$} & \multicolumn{1}{c|}{$\ld^{N-2}\lmax^{4-N}(\lmax/\ld^{N-2})^{1-2\Delta}$}\\
\hline
\end{tabular}
\caption{Scaling of the $(S/N)^2$ in the squeezed limit for $N$-point correlators, when the damping effect is dominant.}
\label{tab:squeezed_limit}
\end{table}

\subsection{The collapsed limit}
\label{subsec:collapsed_limit}
Following the same steps of Sec.~\ref{subsec:squeezed_limit}, in this section we derive the theoretical estimations for the collapsed limit of $N>3$ correlators. We show that, for certain values of $\Delta$, the collapsed limit provides a better scaling than mode-counting, for both 3D and 2D estimates. 

Using the same notation as Sec.~\ref{sec:ng_from_inflation}, we denote the ``left" momenta $\{k_1,\cdots k_M\}\sim k_L$ and the ``right" momenta $\{k_{M+1},\cdots k_N\}\sim k_R$, with $M>1$. In this limit, the diagonal $k_I\equiv k_1+\cdots +k_M$ is much smaller than $k_L,k_R$. Then, replacing the $(N-1)$-spectrum defined in Eq.~\eqref{eq:n_point_collapsed} into the volume survey $(S/N)^2$, we obtain
\begin{equation}
\label{eq:snr_collapsed_computation}
\begin{split}
    \left(\frac{S}{N}\right)^2_{(N)}\sim\int_{\kmin}^{c \kmax}\mathrm{d}^3 k_I\int_{c\kmax}^{ \kmax}\mathrm{d}^{3A} k_R\mathrm{d}^{3B} k_\mathrm{L}\,\frac{k_R^{-6A-2\Delta}k_L^{-6B-2\Delta}k_I^{-6+4\Delta}}{k_R^{-3(A+1)}k_L^{-3(B+1)}}\sim \kmin^{4\Delta-3}\kmax^{6-4\Delta},
\end{split}
\end{equation}
where $A = N-M-1$ and $B = M-1$. We find that for values $\Delta<3/4$ the scaling increases with respect to mode-counting as
\begin{equation}
    \left( \frac{S}{N}\right)^2_{(N)}\sim\kmax^3\left(\frac{\kmax}{\kmin}\right)^{3-4\Delta},\quad \Delta<3/4.
\end{equation}
Along this lines, given the collapsed angular $(N-1)$-spectrum
\begin{equation}
   \lim_{\ell_I \ll \ell_L, \ell_R} F_\zeta(\ell_I, \ell_L, \ell_R)\propto \frac{1}{\ell_R^{2A}\ell^{2B}_L \ell^2_I } \left(\frac{\ell_I^2}{\ell_R \ell_L}\right)^{2\Delta/3},
    \label{eq:cmb_n_point_collapsed}
\end{equation}
we obtain $(S/N)_{(N)}^2\sim \lmin^{8\Delta/3-2}\lmax^{4-8\Delta/3}$.
As for the 3D survey, $\Delta=3/4$ represents a threshold values below which the scaling improves as
\begin{equation}
    \left( \frac{S}{N}\right)^2_{(N)}\sim\lmax^2\left(\frac{\lmax}{\lmin}\right)^{2-8\Delta/3}, \quad \Delta<3/4.
\end{equation}
As a result, even if our estimates are highly idealized and optimistic, collapsed $(N-1)$-spectra might be very sensitive to non-Gaussianity arisen from a light particle driven inflationary process. In particular, local type of non-Gaussianity produces a $\lmax^4$ scaling ($\kmax^6$ for a 3D survey), as was found in Refs.~\cite{kogo:angular_trispectrum,bordin:higher_spin_cmb_statistics}. We summarize the results in Tab.~\ref{tab:3d2d_undamped}. 
\begin{table}[h!]
\renewcommand{\arraystretch}{1.2} 
\centering
\begin{tabular}{ccc}
& \multicolumn{2}{c}{Collapsed limit $\ell\ll \ld$}  \\ \cline{2-3}
\multicolumn{1}{c|}{}  & \multicolumn{1}{c|}{3D} & \multicolumn{1}{c|}{2D ($\ell\ll \ld$)} \\
\cline{2-3}
\noalign{\vskip\doublerulesep{}
         \vskip-\arrayrulewidth}
\hline
\multicolumn{1}{|c|}{$\Delta< 3/4$} & \multicolumn{1}{c|}{$\kmax^3(\kmax/\kmin)^{3-4\Delta}$} & \multicolumn{1}{c|}{$\lmax^2(\lmax/\lmin)^{2-8\Delta/3}$} \\ \hline
\multicolumn{1}{|c|}{$\Delta = 3/4$} & \multicolumn{1}{c|}{$\kmax^3\log(\kmax/\kmin)$} & \multicolumn{1}{c|}{$\lmax^{2}\log(\lmax/\lmin)$}  \\ \hline 
\multicolumn{1}{|c|}{$\Delta > 3/4$} & \multicolumn{1}{c|}{$\kmax^3$} & \multicolumn{1}{c|}{$\lmax^2$}\\
\hline
\end{tabular}
\caption{General scaling behavior in the collapsed limit for the $(S/N)^2$ in 3D and 2D for $N$-point correlation functions.}
\label{tab:3d2d_undamped}
\end{table}

Given the enhanced scaling in the collapsed limit for some values of $\Delta$, and in analogy with the squeezed bispectra, we might expect these shapes to be only mildly affected by damping in the CMB. 
As an example, let us consider the angular damped trispectrum in the collapsed limit
\begin{equation}
    T(\ell_1,\ell_2,\ell_3,\ell_4)\sim \frac{\ld^3}{(\ell_1 \ell_3 \ell_I)^{3}} \left(\frac{\ell_I^{2}}{\ell_1 \ell_3}\right)^{\Delta}.
    \label{eq:cmbdamped_trispectrum_collapsed}
\end{equation}
In this case, we find $(S/N)_{(4)}^2\sim \ld^2(\lmax/\ld)^{4(1-\Delta)}$, therefore trispectra shapes in the interval $0<\Delta<1$ have a favorable scaling despite the effect of damping. The fact that for $\Delta = 0$ yields a scaling of $\lmax^4$, as in the $\ell\ll\ld$ scenario, indicates that damping has little influence on such highly collapsed trispectra.

These results can easily be extended to general collapsed $N$-point correlation function
\begin{equation}
   \lim_{\ell_I \ll \ell_L, \ell_R} F_\zeta(\ell_I, \ell_L, \ell_R)\propto \frac{\ld^{N-1}}{\ell_R^{3A}\ell^{3B}_L \ell^3_I } \left(\frac{\ell_I^2}{\ell_R \ell_L}\right)^{\Delta}.
    \label{eq:cmbdamped_n_point_collapsed}
\end{equation}
Tab.~\ref{tab:collapsed_limit} summarizes the results. Similar to having nested squeezed limits, there is the possibility of nested collapsed limits, which we investigate in App.~\ref{app:multiple_squeezed_collapsed}. In the single collapsed limit scenario, for a massless particle ($\Delta=0$) there is still potential to gain signal-to-noise when moving towards higher resolution (higher $\lmax$), as long as $N<8$, whereas in the general case this enhancement is dependent on the specific number of nested collapsed limits. In the section below, we discuss in more depth the extent of these optimistic scalings in the specific example of the trispectrum.

\begin{table}[h!]
\renewcommand{\arraystretch}{1.2} 
\centering
\begin{tabular}{ccc}
& \multicolumn{2}{c}{Collapsed limit $\ell\gg\ld$}  \\ \cline{2-3}
\multicolumn{1}{c|}{}  & \multicolumn{1}{c|}{Trispectrum} & \multicolumn{1}{c|}{$(N-1)$-spectrum} \\
\cline{2-3}
\noalign{\vskip\doublerulesep{}
         \vskip-\arrayrulewidth}
\hline
\multicolumn{1}{|c|}{$\Delta<1$} & \multicolumn{1}{c|}{$\ld^2(\lmax/\ld^2)^{4(1-\Delta)}$} & \multicolumn{1}{c|}{$\ld^{N-2}\lmax^{4-N}(\lmax/\ld^{N-2})^{4(1-\Delta)}$} \\ \hline
\multicolumn{1}{|c|}{$\Delta = 1$} & \multicolumn{1}{c|}{$\ld^{2}\log(\lmax/\ld)$} & \multicolumn{1}{c|}{$\ld^{2}\lmax^{4-N}\log(\lmax/\ld)$} \\ \hline 
\multicolumn{1}{|c|}{$\Delta > 1$} & \multicolumn{1}{c|}{$\ld^2(\lmax/\ld^2)^{4(1-\Delta)}$} & \multicolumn{1}{c|}{$\ld^{N-2}\lmax^{4-N}(\lmax/\ld^{N-2})^{4(1-\Delta)}$}\\
\hline
\end{tabular}
\caption{The table shows the scaling estimations for 2D in the collapsed limit when the damping effect is dominant.}
\label{tab:collapsed_limit}
\end{table}

\subsubsection{A comment on the enhanced scaling}\label{subsec:CollapsedScaling}

The improved scaling of the $(N-1)$-spectra for $N>3$ in the (nested) collapsed limit immediately raises the question whether trispectrum (and beyond) measurements can provide more precise constraints on the primordial interaction couplings. Here we will limit the discussion to the CMB trispectrum, compared to the bispectrum, but the arguments can be applied more generally to $(N-1)$-spectra, including nested collapsed limits, and to LSS surveys in 3D. 
Earlier work~\cite{kogo:angular_trispectrum} had noticed the enhanced trispectrum scaling for local type non-Gaussianity and the question of whether the trispectrum can better constrain $f_{\mathrm{NL}}$ within that model has been extensively examined in Ref.~\cite{creminelli:estimators_local_ng}. In that work the authors consider the local model, so named as the associated non-Gaussianity can be expressed using the following local modifications to the linear perturbation 
\begin{align}
\Phi(\mathbf{x}) = \Phi_{L}(\mathbf{x})+f_{\mathrm{NL}}( \Phi_{L}^2(\mathbf{x})-\langle  \Phi_{L}^2(\mathbf{x})\rangle) + g_{\mathrm{NL}}\Phi_{L}^3(\mathbf{x}),
\label{eq:fnlfield}
\end{align}
where $ \Phi_{L}(\mathbf{x})$ is the linear, Gaussian perturbation. They show that when the $S/N$ of the trispectrum starts to exceed that of the bispectrum, additional non-Gaussian terms become important, such that the bispectrum signal-to-noise is always larger than that of the trispectrum. Heuristically, this can be understood knowing that in the local model the trispectrum amplitude is $\mathcal{O}(f_{\mathrm{NL}}^2)$ and the non-Gaussian contributions to the trispectrum variance are of the same order. Thus, when the trispectrum signal exceeds the Gaussian variance, so do the non-Gaussian contributions to the variance, completely canceling out enhanced scaling. 

Eq.~\eqref{eq:fnlfield} has two leading contributions when computing the 4-point correlation function. The first contribution is coming from the cubic term $\propto g_{\rm NL}$. By definition, the associated $(S/N)^2\propto \lmax^2$. Dimensionally, we can assume that $g_{\mathrm{NL}}=f_{\mathrm{NL}}^2\alpha$. Therefore the only way for this contribution to be observed with $S/N$ exceeding that of the bispectrum is to have sufficiently large values of $\alpha$ ($\alpha > 1/(f_{\rm NL} \sqrt{A_s})$). The second contribution in the model has favorable scaling in the collapsed limit, i.e. $(S/N)^2\propto \lmax^4$. We generally refer to the amplitude of this contribution with $\tau_\mathrm{NL}$. However, in this model $\tau_\mathrm{NL} \propto f_{\mathrm{NL}}^2$, with no possibility to fine-tune the overall scaling factor ($\alpha = 6$). As a result, even though the $S/N$ could potentially exceed that of the bispectrum, it requires a very large value of $\alpha$ and the scaling of the signal-to-noise of the trispectrum does not improve over simple mode counting. The fact that $\tau_\mathrm{NL} \propto f_{\mathrm{NL}}^2$ also guarantees that a trispectrum will not have a favorable scaling with respect to the signal $f_{\rm NL}$ compared to the bispectrum. In order to exploit the improved scaling the contribution to $\tau_\mathrm{NL}$ should become independent from the value of $f_{\mathrm{NL}}$. A straightforward way to achieve this is by introducing an additional field that couples to the inflaton. 

In Ref.~\cite{kamionkowski:cmb_statistics_cramerrao}, this idea is generalized. The authors point out that even when the bispectrum estimator is an optimal estimator (one which satisfies the Cram\'er-Rao bound and is therefore the minimum possible unbiased estimator) trispectrum measurements can add statistically independent information! This is possible as trispectrum estimators that measure $f_{\mathrm{NL}}^2$ are biased estimators for $f_{\rm NL}$, and bispectrum estimators are suboptimal estimators for $f_{\mathrm{NL}}^2$. Further, they show that trispectra, whose $S/N$ is dominated by a small number of shapes that can be expressed as two triangles (see Fig.~3 in Ref.~\cite{smith:optimal_cmb_trispectrum}),  
are highly correlated with the bispectrum and thus contain minimal new information. 

From these papers we draw two conclusions.  First, there is no general theorem that prevents trispectra (and higher point correlation functions) from providing more constraining power than bispectra on primordial couplings. Second, the $S/N$ computations above, which neglect non-Gaussian contributions, can be too naive, particularly for collapsed $(N-1)$-spectra. For strongly collapsed trispectra, the non-Gaussian contributions can dominate the variance, reducing (or even completely negating) the information in the trispectrum. This conclusion is however model dependent and when the relation between the trispectrum and bispectrum amplitude deviates from the relation demonstrated for the local model, as in Ref.~\cite{bordin:higher_spin_cmb_statistics}, the trispectrum can be highly informative and constraining.

\section{The CMB bi- and tri-spectrum}
\label{sec:numerical_analysis}
In this section we present a numerical analysis of the $S/N$ for the bispectrum and the trispectrum of the temperature and polarization anisotropies. We assume measurements are cosmic variance limited and assume unlensed spectra (see e.g. \cite{coulton:minimizing_gravlens_bispectrum} for the effect of lensing and how to mitigate it). The analysis in this section will use the full radiation transfer functions, which we calculate numerically using the Boltzmann solver \texttt{CAMB}~\cite{lewis:camb}. We aim to confirm the heuristic results of  Sec.~\ref{sec:theory_snr} in the damped regime $\ell\gg\ld$. Tab.~\ref{tab:cosmology} summarizes the flat $\Lambda$CDM cosmology parameters used.

\subsection{Bispectrum}
In this subsection, we aim to compute
\begin{equation}
\label{eq:cmb_snr}
    \left(\frac{S}{N}\right)^2_{(3)} = \sum_{X_i, X'_i} \sum_{\ell_1\leq\ell_2\leq\ell_3\leq \lmax} \frac{h^{\ell_1\ell_2\ell_3}}{f(\ell_1,\ell_2,\ell_3)} b^{X_1 X_2 X_3}_{\ell_1 \ell_2 \ell_3} \left(C_{\ell_1}^{X_1 X'_1}\right)^{-1} \left(C_{\ell_2}^{X_2 X'_2}\right)^{-1} \left(C_{\ell_3}^{X_3 X'_3}\right)^{-1}b^{X'_1 X'_2 X'_3}_{\ell_1 \ell_2 \ell_3},
\end{equation}
up to $\lmax = 5000$. Here $X_i= \{T,E\}$ where $T,\,E$ refer respectively to temperature and $E$-mode polarization. 

The general functional form of the primordial bispectrum can render the computation of the CMB bispectrum challenging. To mitigate this issue, separable templates, which approximate the theoretical predictions and are suitable for data analysis, have been proposed. In this subsection, we will investigate the local~\cite{komatsu:acoustic_signature_cmb_bispectrum}, equilateral~\cite{creminelli:ng_from_wmap} and orthogonal templates~\cite{senatore:ng_single_field}.

The local template has a corresponding value $\Delta = 0$ and is given by
\begin{equation}
\label{eq:local_bispectrum}
B^\mathrm{local}_\zeta (k_1,k_2,k_3) = \frac{6}{5} f_\mathrm{NL}^\mathrm{local}\bigg[P_\zeta (k_1)P_\zeta (k_2)+\text{2 perms.}\bigg].
\end{equation}
Here $P_\zeta (k) = 2\pi^2 A_\mathrm{s} (k/k_\star)^{n_\mathrm{s}-1} k^{-3}$, where $A_\mathrm{s}$ the amplitude of initial fluctuations, $n_\mathrm{s}$ is the scalar spectral index (see e.g. Tab.~\ref{tab:cosmology}) and $k_\star = 0.05\; \mathrm{Mpc^{-1}}$ is the pivot scale. 
This shape is largest in the squeezed limit, i.e. $k_1\ll k_2\sim k_3$, therefore according to our analytical estimates in Sec.~\ref{subsec:squeezed_limit}, we expect the scaling of the $(S/N)^2$ to be proportional to $\lmax^2$.

Equilateral non-Gaussianity can be captured using the following template
\be
B^\mathrm{equil}_\zeta (k_1,k_2,k_3) &=&  \frac{18}{5}f_\mathrm{NL}^\mathrm{equil} \left[-P_\zeta(k_1)P_\zeta(k_2)-\text{2 perms.}-2P^{2/3}_\zeta(k_1)P^{2/3}_\zeta(k_2)P^{2/3}_\zeta (k_3)+ \right. \nonumber\\ && \left. P^{1/3}_\zeta (k_1)P^{2/3}_\zeta (k_2)P_\zeta (k_3)+\text{5 perms.}\right],
\label{eq:equil_bispectrum}
\ee
and peaks when $k_1 = k_2 = k_3$. It is straightforward to check that this template has $\Delta= 2$, and we expect $(S/N)^2$ to be proportional to $\lmax$. 

A third shape, orthogonal to the equilateral template, 
was introduced in Ref.~\cite{senatore:ng_single_field} and it is parametrized by
\be
B^\mathrm{ortho}_\zeta (k_1,k_2,k_3) &=& \frac{18}{5}f_\mathrm{NL}^\mathrm{ortho} \left[-3 P_\zeta(k_1)P_\zeta(k_2)-\text{2 perms.}-8P^{2/3}_\zeta(k_1)P^{2/3}_\zeta(k_2)P^{2/3}_\zeta(k_3)+ \right. \nonumber\\ && \left. 3P^{1/3}_\zeta (k_1)P^{2/3}_\zeta(k_2)P_\zeta(k_3)+\text{5 perms.}\right].
\label{eq:ortho_bispectrum}
\ee
This shape peaks in both equilateral and flattened configurations, i.e. $k_1 = k_2+k_3$. The orthogonal template above has $\Delta = 1$, however as shown in Ref.~\cite{senatore:ng_single_field}, Eq.~\eqref{eq:ortho_bispectrum} is an approximation of a more numerically challenging shape and does not have the correct scaling in the squeezed limit ($\Delta = 2$). For our purposes however it suffices since it nicely completes a set of templates with $\Delta = 0$, 1, and 2. Our analytical estimate suggests that $(S/N)^2$ should be proportional $\ell_{\rm max}$ because $\Delta>1/2$. Our numerical analysis however will show that the orthogonal template produces an improved scaling, at least over the range we explored ($\lmax = 5000$). We will comment on this discrepancy in some depth below. 

\begin{figure*}[htbp!]
\centering
\includegraphics[width=1.0\textwidth]{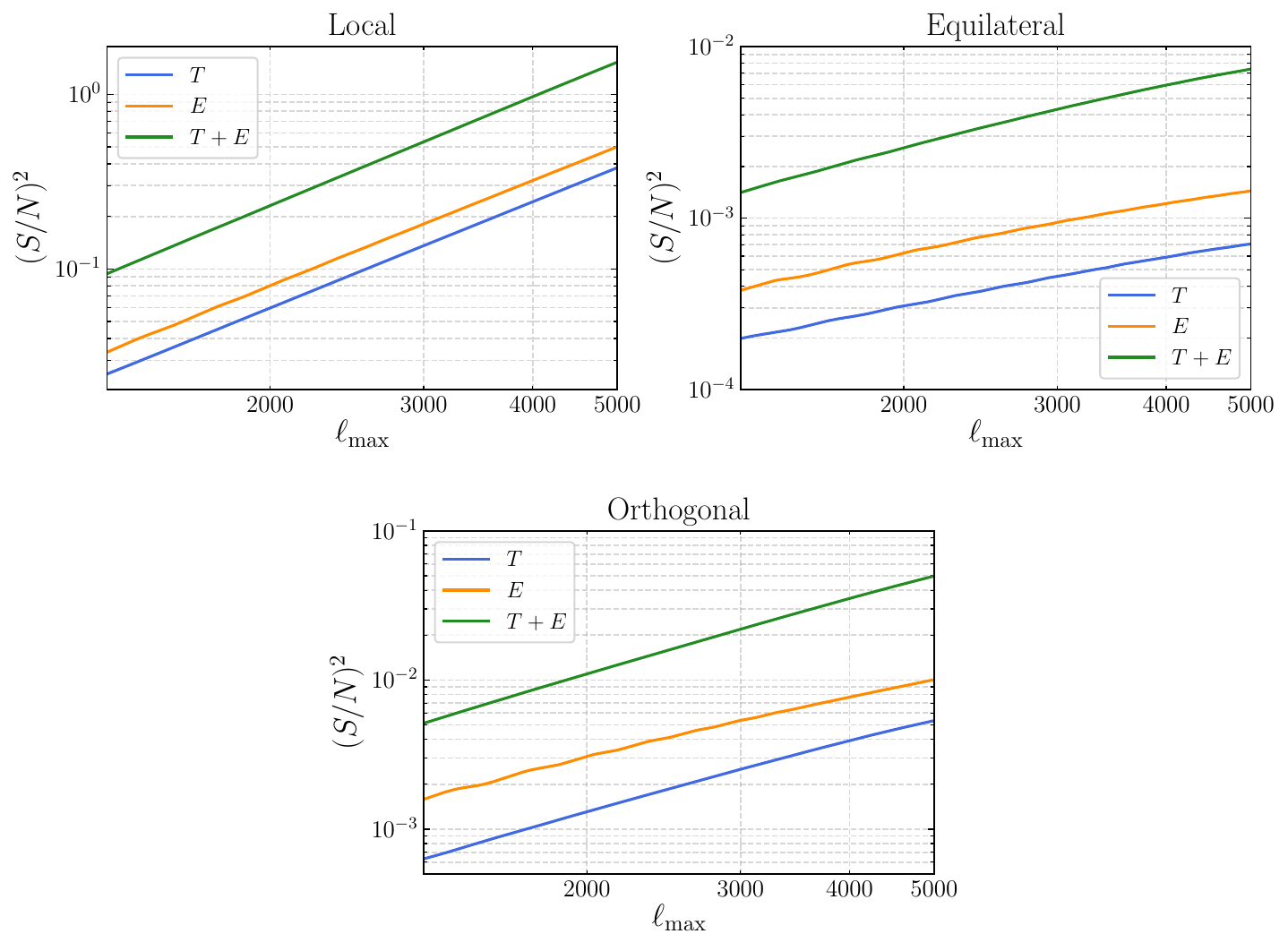}
\caption{Numerical solution of the bispectrum $(S/N)^2$ given in Eq.~\eqref{eq:cmb_snr} in the range $\ld \leq \lmax \leq 5000$, where $\ld \simeq 1300$. \textit{Upper left panel:} $(S/N)^2$ for the local shape for temperature, $E$-mode polarization and in combination. \textit{Upper right and lower panel}:  $(S/N)^2$ for the equilateral and orthogonal shapes respectively. Curves shown are for $f_{\rm NL} = 1$.}
\label{fig:fisher_T_E}
\end{figure*}

In Fig.~\ref{fig:fisher_T_E}, we show the $(S/N)^2$ for \textit{only} temperature, \textit{only} $E$-mode polarization, and \textit{both} temperature and mode polarization anisotropies.

Guided by our analytical results in Sec.~\ref{sec:theory_snr}, we expect the $(S/N)^2$ to be proportional to $\lmax^p$. 
Numerically, we can extract the exponent by taking $\log$-derivative of the $(S/N)^2$, i.e., 
\begin{equation}
\label{eq:scaling_p}
    p = \frac{\mathrm{d}}{\mathrm{d}\log\lmax} \log{[(S/N)^2]}\equiv \frac{\lmax}{(S/N)^2} \frac{\mathrm{d}(S/N)^2}{\mathrm{d}\lmax}.
\end{equation}
We are mostly interested in $p$ in the damped regime, so it suffices to fit the numerical exponent with a constant based on the theoretical scaling in the range $\lmax>\ld$, where $\ld \simeq 1300$. In Tab.~\ref{tab:e_t_scaling} we report our results together with the one standard deviation error.
\begin{table}[t!]
\renewcommand{\arraystretch}{1.2} 
\centering
\begin{tabular}{lccc}
   & \multicolumn{3}{c}{Numerical Scaling Bispectrum}  \\ \cline{2-4} 
\multicolumn{1}{l|}{}                       & \multicolumn{1}{c|}{Local} & \multicolumn{1}{c|}{Equilateral} & \multicolumn{1}{c|}{Orthogonal} \\
\cline{2-4} 
\noalign{\vskip\doublerulesep{}
         \vskip-\arrayrulewidth}
\hline
\multicolumn{1}{|c|}{$T$} & \multicolumn{1}{c|}{$2.0169\pm 0.0003$} & \multicolumn{1}{c|}{$0.912\pm 0.003$}       & \multicolumn{1}{c|}{$1.544\pm 0.002$}      \\ \hline
\multicolumn{1}{|c|}{$E$} & \multicolumn{1}{c|}{$2.005\pm 0.002$} & \multicolumn{1}{c|}{$0.936\pm 0.004$}       & \multicolumn{1}{c|}{$1.326\pm 0.004$}      \\ \hline
\multicolumn{1}{|l|}{$ T+E$} & \multicolumn{1}{c|}{$2.0650\pm 0.0004$} & \multicolumn{1}{c|}{$1.170\pm 0.003$}       & \multicolumn{1}{c|}{$1.658\pm 0.001$}      \\ \hline
\end{tabular}
\caption{Values of the numerical scaling $p$ inferred from Eq.~\eqref{eq:scaling_p} using a least square fit and the standard deviation error. The numerical derivatives are derived in the range $\ld \leq \lmax \leq 5000$. These values are to be compared to the theoretical ones in Tab.~\ref{tab:squeezed_limit}.}
\label{tab:e_t_scaling}
\end{table}
The numerical exponents are in good agreement with the theory estimates in Tab.~\ref{tab:squeezed_limit}, except for the orthogonal shape which shows a better scaling than the expectation in this range of multipoles. While we do not have an analytical argument to explain this behavior, we can derive some intuition from computing the derivative of the scaling defined in Eq.~\eqref{eq:scaling_p}. We find that, while both equilateral and orthogonal shapes yield a negative derivative, the magnitudes of the derivatives suggest that the orthogonal shape converges more slowly towards the limiting scaling\footnote{The flat shape has a similar slow convergence behavior, having $\Delta=1$. See e.g.~\cite{Fergusson_2009} Eq.~(21).}.
This could suggest that the scaling predicted by our analytical analysis is not yet captured when limiting our numerical analysis to $\lmax \leq 5000$. Another explanation could be that our heuristic analytical derivation does not capture the details of the full radiative transfer function, resulting in intermediate scalings for $\Delta$ close to the critical value $\Delta_c$. The main point however, as confirmed by our numerical analysis, is that for $\Delta > \Delta_c$ (equilateral-like shapes) the scaling is affected much more severely by damping then for shapes where $\Delta < \Delta_c$ (local-like shapes). 

Our analytical calculations were strictly derived as an approximation of the temperature transfer functions (see App.~\ref{app:damping_cmb}), but polarization fluctuations are similarly affected by the thickness of last scattering, evident from our numerical calculations. Combining the temperature and polarization signatures can lead to improvement of the scaling as can be read off from Tab.~\ref{tab:e_t_scaling} for the orthogonal and equilateral shapes.  
While we do not have a full qualitative derivation why this happens, we can make the following heuristic argument. The reduced scaling for shapes with $\Delta > \Delta_c$ is caused by the blurring of the modes on small scales. At the level of the radiative transfer function, the temperature and polarization modes oscillate (around zero) and are out of phase. When combining polarization and temperature measurements, where the zero points of each individual transfer function yielded zero response and a loss of primordial signal, the product of these transfer functions can now limit some of these losses around the zero points of the transfer function (the zero points are recovered only in the auto correlation of the fields). The net effect is a `de-blurring' of the last scattering surface. While this leads to immediate improvements on the scaling of spectra with $\Delta > \Delta_c$, since blurring does very little to spectra with $\Delta < \Delta_c$, no change in the scaling is anticipated. Hence the scaling of the $S/N$ for e.g. the local shape does not benefit from adding polarization. 

{\paragraph{Lensing effect on the $S/N$} In this work we neglect the effect of gravitational lensing on small scales, assuming that our CMB fields are unlensed. Lensing is a second order effect that generates large non-Gaussianity, which introduces corrections to the statistical properties of the CMB. 

If not accounted for, lensing would affect the measurements of non-Gaussianities in the CMB in two distinct ways. First, it introduces a signal (e.g. the ISW-lensing bispectrum, see \cite{goldberg:isw_lensing}) which can interfere with the primordial signal. Second, it will introduce extra covariance~\cite{kayo:info_content_weak_lensing,coulton:secondary_ng_actpol_planck}. In CMB analysis one typically removes or marginalizes the additional signal (for example, this was done in the Planck analysis~\cite{planck:lensing}). On the other hand, the extra covariance is a concern only for post-Planck analyses and, if not accounted for, it would affect the scaling derived in this paper.

On small scales, lensing will dominate over the primary modes in the power spectrum, especially for $E$ mode polarization.
As a consequence, even if we assume a Gaussian covariance, we should anticipate loss of signal-to-noise over the no-lensing derivation presented here. Furthermore, the non-Gaussian nature of lensing excites all higher order moments in the covariance. Of the shapes we consider here, the local one is the most affected, while the impact on the equilateral and orthogonal shapes is lower but still non-negligible. Recently, this effect and the techniques to mitigate it have been extensively discussed in Ref.~\cite{coulton:minimizing_gravlens_bispectrum} for the CMB bispectrum. The good news is that it was shown that by delensing the data before estimating the signal, almost all extra covariance is removed. As a bonus, signal biases are also removed when the data is delensed. This suggests that, neglecting all else, the scalings derived in this paper for the bispectrum should hold.

We expect lensing will also affect the $S/N$ of higher $N$-point correlators, both for squeezed and collapsed shapes. In this case, the higher number of fields involved makes the delensing procedure more challenging, as the noise bias terms grow in number and complexity. The noise biases are introduced when you apply the $N$-point estimator on (partially) delensed data using that same fields you use to estimate the non-Gaussian statistics (see appendix of Ref.~\cite{coulton:minimizing_gravlens_bispectrum} for an explicit calculation). It was shown that this does not lead to biases at lowest order in the bispectrum due to the odd number of fields. However, it is likely that such procedure would lead to noise biases in even spectra. It would be interesting to investigate in depth the size of these biases. In principle they can be computed and subtracted, similarly to how noises biases are removed in lensing potential reconstruction. A potential strategy to avoid these biases would be to reconstruct the lensing potential with an external tracer such as the Cosmic Infrared Background. While not perfect, this should be able to remove a large fraction of the lensing and thus would help to maintain the estimated scalings here.   
Since the focus of the paper is to investigate the information content of CMB primary anisotropies, we leave the generalization of $N$-point correlators of lensed fields to future work.}

\subsection{Trispectrum}
In this subsection, we will perform a numerical Fisher analysis for the trispectrum. We will focus on the separable form of the local primordial trispectrum~\cite{kogo:angular_trispectrum,regan:trispectrum_estimation,smith:optimal_cmb_trispectrum}\footnote{For a semi-analytic method to compute cosmological angular trispectra related to the discussion on cosmological correlators in Sec.~\ref{sec:ng_from_inflation}, see e.g.~\cite{lee:trispectra}.}
\begin{equation}
\label{eq:local_trispectrum_AB}
T^{\local}_{\zeta}(\bsk_1,\bsk_2,\bsk_3,\bsk_4,\boldsymbol{K})=T^{\tau_\mathrm{NL}}_{\zeta}(\bsk_1,\bsk_2,\bsk_3,\bsk_4,\boldsymbol{K})+ T^{g_\mathrm{NL}}_{\zeta} (\bsk_1,\bsk_2,\bsk_3,\bsk_4,\boldsymbol{K}),
\end{equation}
where
\be
\label{eq:local_trispectrum_A}
    T^{\tau_\mathrm{NL}}_\zeta(\bsk_1,\bsk_2,\bsk_3,\bsk_4,\boldsymbol{K}) &=& \tau_\mathrm{NL}[P_\zeta (K) P_\zeta (k_1)P_\zeta (k_3)+\text{11 perms.}],\\
\label{eq:local_trispectrum_B}
    T^{g_\mathrm{NL}}_{\zeta}(\bsk_1,\bsk_2,\bsk_3,\bsk_4,\boldsymbol{K}) &=&\frac{54}{25} g^\local_\mathrm{NL}[P_\zeta (k_2) P_\zeta (k_3)P_\zeta (k_4)+P_\zeta (k_1) P_\zeta (k_2)P_\zeta (k_4)+\text{2 perms.}], \nonumber \\
\ee
with $\tau_\mathrm{NL} \geq (6/5 f^\local_\mathrm{NL})^2$~\cite{suyama:ng_modulated_reheating}. Both the $\tau_{\rm NL}$ and the $g_{\rm NL}$ templates arise from local type of non-Gaussianity. However, here we consider them separately because our analytical analysis suggest these trispectra should have a different scaling with $\lmax$. The $\tau_\mathrm{NL}$-trispectrum is largest in the collapsed limit with $\Delta = 0$, and we expect $(S/N)^2\propto \lmax^4$. On the other hand, as shown in Sec.~\ref{subsec:squeezed_limit}, the $g_\mathrm{NL}$-trispectrum is largest in the double squeezed limit, producing a scaling of $(S/N)^2\propto \lmax^2$.

\begin{figure}[h!]
    \centering
    \includegraphics[width=1.0\textwidth]{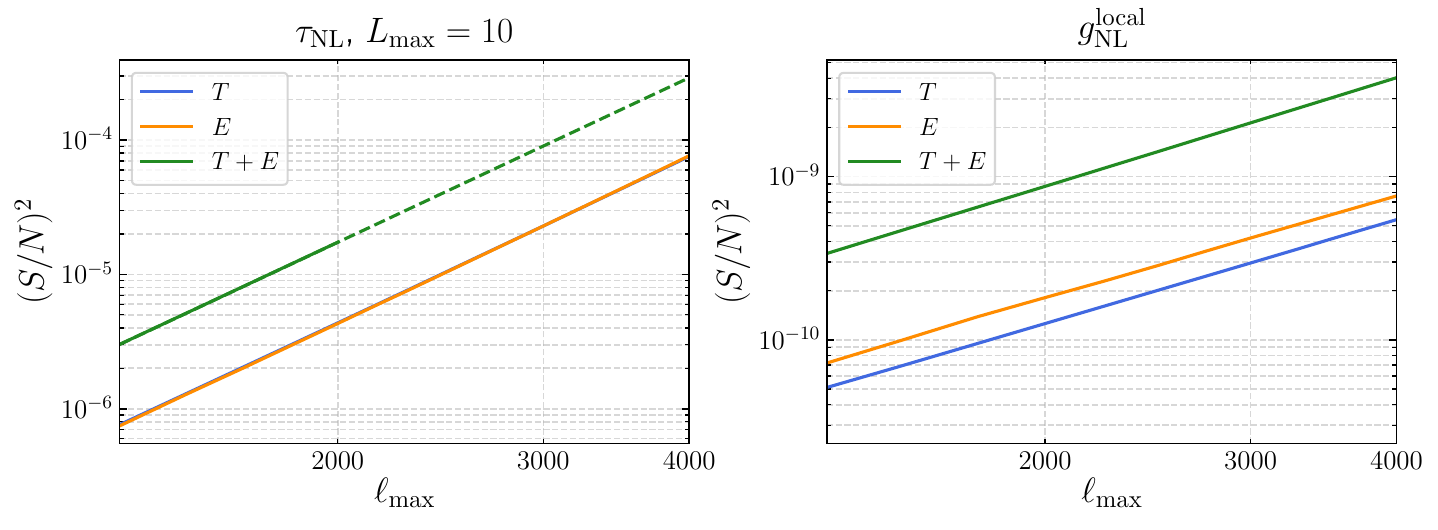}
    \caption{Numerical solution of the trispectrum $(S/N)^2$, given in Eq.~\eqref{eq:cmb_trispectrum_snr} in the range $\ld \leq \lmax \leq 4000$, where $\ld \simeq 1300$. \textit{Left panel}: $(S/N)^2$ for the $\tau_\mathrm{NL}$-trispectrum shape for temperature, $E$-mode polarization and in combination. The $T+E$ solution is shown in the range $\ld \leq \lmax \leq 2000$ and the dashed curve refers to the expected solution at higher multipoles. Notice that the $T$ and $E$ show identical contribution to the $S/N$. This is expected because for the trispectrum in the collapsed limit the two triangles in the limit of small $L$ will be squeezed and their contribution to the trispectrum is expected to be proportional to $\prod _i^4C^X_{\ell_i}$. This cancels exactly with the cosmic variance noise if considering only $T$ or $E$. \textit{Right panel}:  $(S/N)^2$ $g_\mathrm{NL}$-trispectrum shape. Curves shown are for $\tau_{\rm NL} = 1$ and $g_{\rm NL} = 1$.}
    \label{fig:tauNL_gNL}
\end{figure}

We compute the $S/N$ 
\be
\left(\frac{S}{N}\right)^2_{(4)} &=& \sum_{X_i, X'_i}\sum \limits_{\ell_i} \sum \limits_L \frac{1}{(2L+1)} T^{X_1 X_2 X_3 X_4}_{c,\ell_1\ell_2\ell_3\ell_4}(L)\left(\mathbb{C}^{-1} \right)_{\ell_1}^{X_1 X'_1}\left(\mathbb{C}^{-1} \right)_{\ell_2}^{X_2 X'_2}\times  \nonumber \\
&& \left(\mathbb{C}^{-1} \right)_{\ell_3}^{X_3 X'_3} \left(\mathbb{C}^{-1} \right)_{\ell_4}^{X_4 X'_4}T^{X'_1 X'_2 X'_3 X'_4}_{c,\ell_1\ell_2\ell_3\ell_4}(L),
\ee
for the angular trispectra in Eq.~\eqref{eq:local_trispectrum_A} and Eq.~\eqref{eq:local_trispectrum_B}. The computational requirements are challenging as in general the above equation scales as $\mathcal{O}(r^2 \ell_{\rm max}^4)$. With the reformulation and approximations described in App.~\ref{app:cmb_fullsky} we show that we can evaluate $(S/N)^2_{\tau_{\mathrm{NL}}}$ with $\mathcal{O}(r\ell_{\rm max}^4)$ and  $(S/N)^2_{g_{\mathrm{NL}}}$ with $\mathcal{O}(r^2\ell_{\rm max}^3)$ computations. For this reason we limit ourselves to $\lmax = 4000$. When combining temperature and polarization, the number of computations increases by a factor of 16 for $\tau_{\rm NL}$ and 2 for $g_{\rm NL}$. We therefore limit our analysis of the combined temperature and polarization modes to $\lmax = 2000$ and  $\lmax = 4000$ for $\tau_{\mathrm{NL}}$ and $g_{\rm NL}$ respectively. The $g_{\mathrm{NL}}$-angular trispectrum does not explicitly dependent on the diagonal mode $L$, while the $\tau_\mathrm{NL}$-angular trispectrum does and we compute $S/N$ for the $\tau_\mathrm{NL}$-angular trispectrum up to $L_{\rm max}=10$. We varied $L_{\rm max}$ and found it did not change our results significantly, which is in accordance with the findings in Ref.~\cite{kogo:angular_trispectrum}. Further details of the computation can be found in App.~\ref{app:cmb_fullsky}. The results are shown in Fig.~\ref{fig:tauNL_gNL}.

We repeat the analysis done for the bispectrum and extract the scaling of the $(S/N)^2$ using Eq.~\eqref{eq:scaling_p} in the damped regime $\lmax>\ld$. The results are summarized in Tab.~\ref{tab:e_t_scaling}. We find a scaling for the $(S/N)^2_{\tau_{\rm NL}} \propto \lmax^4$ in accordance with our analytical forecast. When combining temperature and polarization, the scaling gets a little worse, but the change is caused by the limited range in $\lmax$ over which the scaling is determined. If we limit ourselves in the same range in temperature, we obtain a consistent scaling. Our analysis shows that the scaling of $(S/N)^2_{\tau_{\rm NL}}$ is saturated for the damped CMB, and the scaling can not be improved when combining temperature and polarization. This confirms the results we found for the bispectrum, where the local bispectrum does not appear to benefit when combining the data, because damping has a little effect on the scaling. Tab.~\ref{tab:3d2d_undamped} suggests that there is no difference between the damped an undamped tracers in the collapsed limit when considering a shape with $\Delta = 0$ for the trispectrum. 
\begin{table}[h!]
\renewcommand{\arraystretch}{1.2} 
\centering
\begin{tabular}{lccc}
   & \multicolumn{2}{c}{Numerical Scaling Trispectrum}  \\ \cline{2-3} 
\multicolumn{1}{l|}{}                       & \multicolumn{1}{c|}{Local $\tau_\mathrm{NL}$} & \multicolumn{1}{c|}{Local $g_\mathrm{NL}$} \\
\cline{2-3} 
\noalign{\vskip\doublerulesep{}
         \vskip-\arrayrulewidth}
\hline
\multicolumn{1}{|c|}{$T$} & \multicolumn{1}{c|}{$4.105\pm 0.001$} & \multicolumn{1}{c|}{$2.16 \pm 0.07$}         \\ \hline
\multicolumn{1}{|c|}{$E$} & \multicolumn{1}{c|}{$4.136\pm 0.002$} & \multicolumn{1}{c|}{$2.10\pm 0.05 $}           \\ \hline
\multicolumn{1}{|l|}{$ T+E$} & \multicolumn{1}{c|}{$4.070\pm 0.001$} & \multicolumn{1}{c|}{$2.23 \pm 0.05$}          \\ \hline
\end{tabular}
\caption{Values of the numerical scaling $p$ inferred as in Eq.~\eqref{eq:scaling_p} using a least square fit, together with the standard deviation error. The numerical derivatives are derived in the range $\ld \leq \lmax \leq 4000$, with the exception of combined temperature and polarization data for the $\tau_\mathrm{NL}$-trispectrum which is computed in $\ld \leq \lmax \leq 2000$.
The $T+E$ scaling for the $\tau_\mathrm{NL}$-trispectrum is lower that the temperature one because the range considered is smaller. Indeed, when considering $\ld \leq \lmax \leq 2000$, we find the temperature $(S/N)^2$ has $p = 4.058$. The error on the $g_\mathrm{NL}$-trispectrum is large due to the few data points used to infer the scaling.}
\label{tab:e_t_scaling_trispectrum}
\end{table}

For $g_{\rm NL}$ we obtain a scaling close to $(S/N)^2_{g_{\rm NL}} \propto \lmax^2$. The numerical scaling is a little higher than the scaling predicted by our analytical analysis, but has a fairly large std, because of the low number of samples we were able to compute. The scaling is not expected to benefit much from adding temperature and polarization measurements as we do not expect the scaling of the $S/N$ of the squeezed bispectrum to exceed the scaling of the squeezed trispectrum. Hence, at best for an undamped CMB we anticipate $(S/N)^2_{g_{\rm NL}} \propto \lmax^2 \log \lmax/\lmin$. 

\section{Discussion and conclusions}
\label{sec:discussion_conclusion}
In this paper we built a quantitative and intuitive understanding of scaling (in inverse units of resolution) of the signal-to-noise of primordial $(N-1)$-spectra. Statistical fluctuations in the field that sources the observed density and radiation fluctuations in the universe have the potential to provide  evidence for the dynamics of inflation and  allow us to probe new physics through relic signatures of new particles that modify these $(N-1)$-spectra. From a theoretical point of view, future constraints on these spectra and/or the detection of any of these signatures would be extremely valuable, on par with the discovery of the constituent particles of the standard model, like the Higgs. It is therefore of immediate relevance how we expect future constraints to improve as a function of resolution. In this work we considered the ideal limits in the noiseless case, while assuming that there are no other sources of non-Gaussianity that could hinder a detection (either via signal confusion or by introducing extra variance). Hence all results presented in this paper should be considered in this context, i.e. they present an ideal scenario.  With more realistic assumptions, the scalings derived are likely to get worse and constraints and/or detections will become more challenging. At the same time, we also limit our analytical estimates to just a single tracer field, e.g. in the case of the CMB we only consider the CMB temperature field. As shown in our numerical forecasts, it turns out that adding polarization improves scalings in the CMB, which would mean that our analytical findings are pessimistic compared to the case when multiple tracers are combined. 

Naively, the scaling of the signal-to-noise for all spectra grows proportionally to the highest observed multipole. Our simplified setup gave us insight into how such scalings come about, and how the naive intuition fails. Specifically, we have shown that, due to damping and projection effects, scaling is much poorer, even converging for higher-point correlation functions. Nonetheless, for certain scalings around squeezed and collapsed limits, there is a smaller loss of information when modes are damped. The most important results are summarized in Tab.~\ref{tab:all_scalings}. We have compared our analytic estimates to a full-sky numerical analysis, including the additional signal of polarization. 

Let us summarize the main points of our analysis:

\begin{itemize}
\item The signal-to-noise scaling is reduced due to blurring of the last scattering surface at short distances. This blurring is caused by the combination of projection and damping, but the loss of signal is not due to exponential decay, as both signal and noise are equally damped.

\item The behavior of the $N$-point correlator in the squeezed limit or the collapsed limit for $N>3$ can significantly impact the scaling of the signal-to-noise with the number of modes, even with a reduced range of momenta probing this limit. 

\item Blurring affects equilateral-like $N$-point correlators much more than squeezed ones. 

\item We confirm that our analytical estimates capture the scaling in the limit $\ell \gg \ld$ after comparing these to CMB forecasts using the full radiation transfer equations for the bispectrum and trispectrum. 

\item This comparison also shows that adding polarization data will generally improve the scaling. The improvement is not observed for spectra that are not already close to mode-counting, such as squeezed spectra. 

\item Comparison between analytical and numerical results also shows that while our analytical results seem to suggest a critical value for $\Delta$ above which scaling changes, this cutoff is less clear when applying the full radiative transfer equations. If we compute the derivative of the scaling in Eq.~\eqref{eq:scaling_p}, we find a hierarchy between shapes at low multipoles, depending on the value of $\Delta$, but scaling does converge to the critical scaling at high $\ell_{\rm max}$. 
We conclude that either we have to go to higher $\ell_{\rm max}$ in our full radiative transfer numerical calculations to reach critical scaling, given that our analytical estimates consider scales much larger than the damping scale, or that the full sky radiative transfer functions introduce complexity to the scaling that we can not capture with our simplified estimates for shapes that are near to the critical value $\Delta_c$. 

\item We recover a result that was found earlier that the collapsed limit of the trispectrum can have a signal-to-noise ratio that scales better with resolution than the bispectrum. Our analytical analysis has allowed us to extend this to $(N-1)$-spectra for $N>4$. The enhanced scaling is a function of $\Delta$ (in general) and $N$ (for damped tracers). We have elaborated on this finding in some depth in Sec.~\ref{subsec:CollapsedScaling}. The  enhancement saturates above $N = 4$ even for undamped tracers for a single collapsed limit. However for multiple collapsed limit the enhancement can be larger (see App.~\ref{app:multiple_squeezed_collapsed}).

\item Our analysis assumed that the $S/N$ can be estimated by taking specific kinematic limits, but ignores the remaining angular dependence of the $(N-1)$-spectra. While we empirically find this to be valid for the local and equilateral bispectrum and the local trispectrum, it is possible that our treatment could miss scalings due to phase space limits not captured by the equilateral, collapsed or squeezed limits. For example, the folded limit in the bispectrum $k_i+k_j \sim k_k$ $(i\neq j\neq k)$ could be responsible for the some of the improved scaling we find for the orthogonal template. Keeping explicit angular dependence will complicate the analytical computations and we will leave this for future investigations.
\end{itemize}

 \begin{table}[h!]
 \renewcommand{\arraystretch}{1.2} 
\begin{tabular}{r|c|c|c|l|l}
\cline{2-5}
\multicolumn{1}{l|}{}           & \multicolumn{2}{c|}{2D} & \multirow{2}{*}{3D } & \multicolumn{1}{c|}{\multirow{2}{*}{Notes}} &  \\ \cline{2-3}
\multicolumn{1}{l|}{} & Undamped & Damped &   & \multicolumn{1}{c|}{}                       &  \\ \cline{2-5}
\noalign{\vskip\doublerulesep{}
         \vskip-\arrayrulewidth}
\cline{1-5}
\multicolumn{1}{|r|}{Scaling}   &     $\ell_{\rm max}^2$        &   $ \ell_{\rm max}^{4-N} $       &          $k_{\rm max}^3$            &    For $N=4$ and $\ell\gg\ld$, $(S/N)^2\sim\log \ell_{\rm max}$                                  &  \\ \cline{1-5}
\multicolumn{1}{|r|}{Squeezed}  &    $\ell_{\rm max}^2 $          &   $ \ell_{\rm max}^{5-N-2 \Delta_S}$      &     $k_{\rm max}^3$                &  \begin{tabular}{@{}l@{}} Undamped $\log$-enhancement for $\Delta_S=0$. \\ Damped enhancement for $\Delta_S<1/2$. \end{tabular}                                          &  \\ \cline{1-5}
\multicolumn{1}{|r|}{Collapsed} &   $\ell_{\rm max}^{4-8 \Delta_C/3}$           &     $\ell_{\rm max}^{8-N-4\Delta_C}$     &      $k_{\rm max}^{6-4\Delta_C}$               &                                              \begin{tabular}{@{}l@{}} Undamped enhancement for $\Delta_C<3/4$. \\ Damped enhancement for $\Delta_C<1$. \end{tabular} &  \\ \cline{1-5}
\end{tabular}
\caption{Summary of the theoretical estimations of the $(S/N)^2$ scaling provided in Sec.~\ref{sec:theory_snr}. The analysis does not include nested squeezed or collapsed limits, where the scaling might change, as showed for the $g_\mathrm{NL}$-trispectrum.}
\label{tab:all_scalings}
\end{table}
An important overarching conclusion is that equilateral-like $(N-1)$-spectra are harder to constrain, because there is very little signal in the squeezed and collapsed limits. For an undamped equilateral-like tracer in 3-dimensions, mode-counting is applicable, but no enhancement is possible in the collapsed limit. If information is damped, the difference between squeezed and equilateral shapes becomes even more pronounced. While this loss of information is guaranteed in the CMB, it is imaginable that something similar could happen in large scale structure measurements. For example, a promising avenue to improve constraints on primordial non-Gaussianity is intensity mapping \cite{Karagiannis2019,Bandura:2019uvb}. Clearly, intensity mapping has a lot of similarities with the CMB, as one can imagine that each tomographic bin forms a last scattering surface (with thickness $\Delta z$) of the photons that make up the intensity signal. For that reason, if the bin is too wide, it is possible that information will be lost due to artificial blurring, and this loss would be most evident in equilateral $(N-1)$-spectra. There is in principle no fundamental limitation on the width of a tomographic bin. But it does suggest that one should carefully consider observational and data analysis strategies to make sure you are not throwing away information when looking for these signatures in the (processed) data. 

The analysis in this paper was limited to only the temperature and $E$-mode polarization signal. However, it is expected that the scaling derived from kinematic limits will not change when $B$-mode polarization is considered in the cosmic variance limit and assuming no lensing. If however you aim to constrain a cosmological correlator which contains at least one tensor mode, and the tracer for those tensor modes is a $T$-mode or $E$-mode (see e.g. \cite{Meerburg:2016ecv,bartolo:chiral_gw,Duivenvoorden:2019ses,shiraishi:tensor_ng}), the scalar modes in those tracers would quickly overtake the tensor contribution and the scaling should saturate on small angular scales.  

Results obtained in this paper are relevant only for weak non-Gaussianity, where we can expand the non-Gaussian field such as in e.g. Eq.~\eqref{eq:fnlfield}. In case the perturbations are strongly non-Gaussian or in cases where the non-Gaussianities become manifest in the tails of the distribution, such as proposed in Ref.~\cite{Flauger:2016idt} and explored in data in Ref.~\cite{M_nchmeyer_2019}, the derived scalings will not apply. 

Currently the best constraints on primordial non-Gaussianity come from the CMB bispectrum. In the near future, several CMB experiments will improve on these constraints, predominantly\footnote{In addition the sensitivity to polarization will improve which will double the number of modes and, as our analysis shows, when combined with temperature measurements can result in non-trivial improvements on shapes with $\Delta_c > 1/2$.} by increasing the spatial resolution, reaching to higher $\ell_{\rm max}$ \cite{abazajian:cmb4_forecast,ade:so_forecast,Abazajian:2019eic}. An often quoted threshold of any type of non-Gaussianity is to reach $\sigma_{f_{\rm NL}} \sim 1$ \cite{Alvarez:2014vva}. Given the current bounds on orthogonal and equilateral non-Gaussianity $\sigma_{f_{\rm NL}} \sim \mathcal{O}(50)$, reaching that threshold will be challenging if not downright impossible\footnote{An estimate suggests that this would require $\ell_{\rm max} > 25000$ for orthogonal and $\ell_{\rm max} > 100000$ for equilateral, requiring a CMB dish of  $\mathcal{O}(100)$ meters in diameter with a focal plane loaded with detectors. This assumes we would have unconstrained access to primary modes, which we know is not the case already for $\ell_{\rm max}\sim $ few thousand. We should note that this extrapolation relies on the orthogonal template of Eq.~\eqref{eq:ortho_bispectrum}. The correct orthogonal shape should have $\Delta = 2$ (see App.~B of Ref.~\cite{senatore:ng_single_field}) and the expected scaling would be similar to the equilateral shape. } with CMB measurements alone. While this is a somewhat pessimistic reading of our analysis, the fact that the bispectrum is not the only measure of non-Gaussianity, and that spectra with $N>3$ can exhibit very favorable scaling with resolution, implies that the CMB can certainly contribute to the search for primordial non-Gaussianity in general. For example, graviton exchange trispectra are qualitatively similar to $\tau_{\rm NL}$-like non-Gaussianity, while the corresponding bispectra are equilateral-like. Of course, this example poses a challenging detection regardless, due to the overall coefficient being Planck suppressed. It would be interesting to consider other examples displaying a similar mismatch of scaling behavior of the collapsed trispectrum vs. squeezed bispectrum. 

For large scale structure, the situation is generally more optimistic if we consider cosmic variance limits. For all $(N-1)$-spectra at least mode counting applies and enhancement is possible in the collapsed limit. At the same time, we are always hindered by loss of information on small scales due to non-linearities which introduce a non-linear scale $k_{\rm NL}$ beyond which primordial information will be hard to extract. Spectra that explicitly couple small to large scale modes, such as the local bispectrum, introduce observational effects in the bias of the large scale structure. These effects will likely allow us to reach the threshold of $\sigma_{f_{\rm NL}} \sim 1$ for these type of non-Gaussianities. 
In the absence of such a coupling, our constraints will rely on measurement of $(N-1)$-spectra (with $N>2$). Ideally a measurement of these $(N-1)$-spectra aims to avoid both limitations of blurring and the non-linear scale, for example by mapping the density field out to very high redshifts, such as possible with 21cm measurements \cite{Munoz:2015eqa,Meerburg:2016zdz}. While many challenges lie ahead in measuring those fluctuations, it is perhaps the only path towards measurement of $\sigma_{f_{\rm NL}} \sim 1$\footnote{Obviously, for $f_{\rm NL} > 1$, a detection will become more likely.} for shapes with $\Delta > \Delta_c$.

\vskip 10 pt
\noindent{\bf Acknowledgments} We are grateful to Daniel Baumann, Anthony Challinor, Thomas Fl\"oss, Daniel Green, Eiichiro Komatsu, Joel Meyers, Giorgio Orlando, Enrico Pajer, Antonio Riotto, Leonardo Senatore and Eva Silverstein for useful discussions. We would like to thank Enrico Pajer for detailed comments on the manuscript. A.K. and P.D.M. acknowledge support from the Netherlands organization for scientific research (NWO) VIDI grant (dossier 639.042.730). The work of G.P. is funded by NWO, and is part of the Delta-ITP consortium. W.R.C. acknowledges support from the UK Science and Technology Facilities Council (grant number ST/N000927/1).

\appendix

\section{Cosmic microwave background: full-sky}
\label{app:cmb_fullsky}
In this appendix we review and summarize existing results in the literature for the full-sky bispectrum and the trispectrum.

The CMB temperature and polarization anisotropies are expressed in terms of the $a^X_{\ell m}$ spherical harmonics coefficients
\begin{equation}
\label{eq:alm_transfer}
    a^X_{\ell m} = 4\pi (-i)^\ell \int \frac{\mathrm{\mathrm{d}^3}k}{(2\pi)^3}\Delta_\mathrm{\ell}^X (k) \zeta (\bsk)Y_{\ell m}(\hat{\bsk}),
\end{equation}
where $\hat{\boldsymbol{n}}$ is a direction in the sky and $X = T, E$ represent the CMB temperature and $E$-mode polarization respectively. The transfer function $\Delta_\mathrm{\ell}^X$ encodes the linear evolution which relates temperature and polarization anisotropies to the primordial curvature perturbations. For our analysis, we obtain the full radiative transfer functions using the publicly available code \texttt{CAMB}~\cite{lewis:camb}.

From the $a^X_{\ell m}$ coefficients, we can define the rotational-invariant angular power spectrum
\begin{equation}
    \label{eq:powerspectrum_cmb_alm}
    \langle a^{X_1}_{\ell_1 m_1}a^{X_2}_{\ell_2 m_2}\rangle = \delta_{\ell_1\ell_2}\delta_{m_1 m_2} C^{X_1 X_2}_{\ell_1},
\end{equation}
with 
\begin{equation}
\label{eq:cmb_power_spectrum}
    C_\ell^{X_1 X_2} = \frac{2}{\pi}\int \mathrm{d}k \, k^2 P_\zeta(k) [\Delta_\mathrm{\ell}^X (k)]^2.
\end{equation}
Likewise, the three-point correlation function can be conveniently expressed in terms of the so-called ``reduced'' bispectrum, which contains the physical information about non-Gaussian sources, and a geometrical factor~\cite{komatsu:acoustic_signature_cmb_bispectrum}
\begin{equation}
\label{eq:bispectrum_cmb_alm}
    \langle a^{X_1}_{\ell_1 m_1}a^{X_2}_{\ell_2 m_2}a^{X_3}_{\ell_3 m_3}\rangle = \mathcal{G}^{\ell_1 \ell_2 \ell_3}_{m_1 m_2 m_3} b^{X_1 X_2 X_3}_{\ell_1\ell_2\ell_3},
\end{equation}
where the Gaunt integral $\mathcal{G}^{\ell_1 \ell_2 \ell_3}_{m_1 m_2 m_3}$ has a known solution given by
\begin{equation}
\begin{split}
     \mathcal{G}^{\ell_1 \ell_2 \ell_3}_{m_1 m_2 m_3} &= \int d\Omega_r Y_{\ell_1 m_1}(\hat{r}) Y_{\ell_2 m_2}(\hat{r}) Y_{\ell_3 m_3}(\hat{r})\\
     &=\sqrt{\frac{(2\ell_1+1)(2\ell_2+1)(2\ell_3+1)}{4\pi}}\left( \begin{matrix} \ell_1 & \ell_2 & \ell_3 \\ 0 & 0 & 0 \end{matrix}\right) \left( \begin{matrix} \ell_1 & \ell_2 & \ell_3 \\ m_1 & m_2 & m_3 \end{matrix}\right) .
\end{split}
\end{equation}
The two matrices are Wigner 3-$j$ symbols. The ``reduced'' bispectrum $b^{X_1 X_2 X_3}_{\ell_1\ell_2\ell_3}$ is connected to the primordial bispectrum $ B_\zeta(k_1,k_2,k_3)$ (see Eq.~\eqref{eq:moments_curvature}) via
\begin{equation}
\label{eq:reduced_bispectrum}
    \begin{split}
        &b^{X_1 X_2 X_3}_{\ell_1\ell_2\ell_3} = \left(\frac{2}{\pi}\right)^3 \int \mathrm{d}r r^2\prod_{i=1}^3\left(\int \mathrm{d}k_i\, k_i^2\Delta^{X_i}_{\ell_i}(k_i)j_{\ell_i}(k_i r)\right) B_\zeta(k_1,k_2,k_3),
    \end{split}
\end{equation}
with $j_\ell(kr)$ the spherical Bessel function which appear through the Rayleigh expansion formula. In principle, given any primordial bispectrum, we can compute the reduced bispectrum using Eq.~\eqref{eq:reduced_bispectrum}. However, it would involve all possible combinations of $\ell$ and $X$, which is computationally expensive (scaling as $\ell^5$).

If $B_\zeta$ is separable, that is, it can be expressed in terms of a product of functions that depend on a single momentum $k_i$, a fast and efficient way to estimate the reduced bispectrum was introduced by the authors of Ref.~\cite{komatsu:ksw_estimator}, the so-called Komatsu-Spergel-Wandelt (KSW) estimators. If we define the radial functions as
\begin{equation}
\label{eq:radial_funcs}
    \begin{split}
        &\alpha_{\ell}^X(r)  \equiv \frac{2}{\pi} \int \mathrm{d}k\,k^2  \Delta^X_{\ell}(k)j_{\ell}(kr), \quad
\beta_{\ell}^X(r)  \equiv \frac{2}{\pi}  \int \mathrm{d}k\,k^2 \Delta^X_{\ell}(k)j_{\ell}(kr)P_{\zeta}(k), \quad\\
& \gamma_{\ell}^X(r)  \equiv \frac{2}{\pi}  \int \mathrm{d}k\,k^2 \Delta^X_{\ell}(k)j_{\ell}(kr)P^{1/3}_{\zeta}(k), \quad
\delta_{\ell}^X(r)  \equiv \frac{2}{\pi}  \int \mathrm{d}k\,k^2 \Delta^X_{\ell}(k)j_{\ell}(kr)P^{2/3}_{\zeta}(k),
    \end{split}
\end{equation}
then, using the templates introduced in Eqs.~\eqref{eq:local_bispectrum},~\eqref{eq:equil_bispectrum} and~\eqref{eq:ortho_bispectrum}, we obtain the reduced bispectra
\begin{equation}
    \label{eq:reduced_bispectrum_local}
    \begin{split}
      &b^{X_1 X_2 X_3,\local}_{\ell_1\ell_2\ell_3} = \frac{6}{5}f^\local_\mathrm{NL}\int \mathrm{d}r\,r^2 \left[\alpha^{X_1}_{\ell_1}(r)\beta^{X_2}_{\ell_2}(r)\beta^{X_3}_{\ell_3}(r) + {\rm 2 \;perms}\right ],
    \end{split}
\end{equation}
\begin{equation}
    \label{eq:reduced_bispectrum_equil}
    \begin{split}
    b^{X_1 X_2 X_3,\equil}_{\ell_1\ell_2\ell_3}=  \frac{18}{5}f^\equil_\mathrm{NL}&\int \mathrm{d}r\,r^2 \bigg[-\alpha^{X_1}_{\ell_1}(r)\beta^{X_2}_{\ell_2}(r)\beta^{X_3}_{\ell_3}(r) - {\rm 2 \;perms}\\ &-2\delta^{X_1}_{\ell_1}(r)\delta^{X_2}_{\ell_2}(r)\delta^{X_3}_{\ell_3}(r)+\gamma^{X_1}_{\ell_1}(r)\delta^{X_2}_{\ell_2}(r)\beta^{X_3}_{\ell_3}(r) + {\rm 5 \;perms}\bigg],
    \end{split}
\end{equation}
\begin{equation}
    \label{eq:reduced_bispectrum_ortho}
    \begin{split}
    b^{X_1 X_2 X_3,\ortho}_{\ell_1\ell_2\ell_3}=  \frac{18}{5}f^\ortho_\mathrm{NL}&\int \mathrm{d}r\,r^2  \bigg[-3\alpha^{X_1}_{\ell_1}(r)\beta^{X_2}_{\ell_2}(r)\beta^{X_3}_{\ell_3}(r) - {\rm 2 \;perms}\\ &-8\delta^{X_1}_{\ell_1}(r)\delta^{X_2}_{\ell_2}(r)\delta^{X_3}_{\ell_3}(r)+3\gamma^{X_1}_{\ell_1}(r)\delta^{X_2}_{\ell_2}(r)\beta^{X_3}_{\ell_3}(r) + {\rm 5 \;perms}\bigg]. 
    \end{split}
\end{equation}

Similarly, the trispectrum is defined as the connected part of the four point correlation function of temperature and polarization anisotropies in Eq.~\eqref{eq:alm_transfer}
\begin{equation}
\label{eq:trispectrum_alm} 
\begin{split}
    \langle a^{X_1}_{\ell_1 m_1}a^{X_2}_{\ell_2 m_2} a^{X_3}_{\ell_3 m_3} a^{X_4}_{\ell_4 m_4}	\rangle_c &= \sum_{L M}(-1)^M \left( \begin{matrix} \ell_1 & \ell_2 & L \\ m_1 & m_2 & -M \end{matrix}\right) \left( \begin{matrix} \ell_3 & \ell_4 & L \\ m_3 & m_4 & M \end{matrix}\right) T^{\ell_1 \ell_2,\boldsymbol{X}}_{\ell_3 \ell_4} (L), 
\end{split}
\end{equation}
where we denote with $\boldsymbol{X} = X_1 X_2 X_3 X_4$ to lighten the notation. The trispectrum generically consists of the connected part, $T_c$, which contains the non-Gaussian signatures, and the unconnected part, $T_G$, which contains only the angular power spectrum and is ignored here. Using permutation symmetry, we may write the connected part as
\begin{equation}
\label{eq:connectTrispec}
    \begin{split}
        {T_c}^{\ell_1\ell_2,\boldsymbol{X}}_{\ell_3\ell_4}(L) = P^{\ell_1\ell_2,\boldsymbol{X}}_{\ell_3\ell_4}(L)+(2L+1) \sum\limits_{L'}  \bigg(& (-1)^{\ell_2+\ell_3}\begin{Bmatrix} \ell_1 & \ell_2 & L \\ \ell_4 & \ell_3 & L' \end{Bmatrix} P^{\ell_1\ell_3,\boldsymbol{X}}_{\ell_2\ell_4}(L') +\\
        &+(-1)^{L+L'}\begin{Bmatrix} \ell_1 & \ell_2 & L \\ \ell_3 & \ell_4 & L' \end{Bmatrix} P^{\ell_1\ell_4,\boldsymbol{X}}_{\ell_3\ell_2}(L')\bigg),
    \end{split}
\end{equation}
where each unique pairing of the multipoles implies 4 permutations 
\begin{equation}
    \begin{split}
        P^{\ell_1\ell_2,\boldsymbol{X}}_{\ell_3\ell_4}(L) =& t^{\ell_1 \ell_2,\boldsymbol{X}}_{\ell_3\ell_4}(L)+(-1)^{2L+\ell_1+\ell_2+\ell_3+\ell_4} t^{\ell_2 \ell_1,\boldsymbol{X}}_{\ell_3\ell_4}(L)+(-1)^{L+\ell_3+\ell_4} t^{\ell_1 \ell_2,\boldsymbol{X}}_{\ell_4\ell_3}(L)+\\
        &+(-1)^{L+\ell_1+\ell_2} t^{\ell_2 \ell_1,\boldsymbol{X}}_{\ell_4\ell_3}(L).
    \end{split}
\end{equation}
Here, the matrix is the Wigner 6-$j$ symbol, which is a summation over the product of four Wigner 3-$j$ symbols, and $t^{\ell_1 \ell_2,\boldsymbol{X}}_{\ell_3\ell_4}(L)$ is the reduced trispectrum, defined as
\begin{equation}
\label{eq:angular_cmb_trispectrum}
\begin{split}
     t^{\ell_1 \ell_2,\boldsymbol{X}}_{\ell_3 \ell_4}(L) =&   \left(\frac{2}{\pi}\right)^5 h_{\ell_1 \ell_2 L} h_{\ell_3 \ell_4 L}\int   \mathrm{d}r_1  \mathrm{d}r_2 r_1^2 r_2^2\int K \,K^2 j_L(K r_1) j_L(K r_2)\times \\
& \times\prod_{i=1}^4\left(\int \mathrm{d}k_i\, k_i j_{\ell_i}(k_i r_i)\Delta^{X_i}_{\ell_i}(k_i)\right)T_\zeta(k_1,k_2, k_3,k_4),
\end{split}
\end{equation}
with
\begin{equation}
    h_{\ell_i \ell_j L} = \sqrt{\frac{(2\ell_i +1)(2\ell_j +1)(2L +1)}{4\pi}}\left(\begin{matrix}\ell_i & \ell_j & L\\
0&0&0\end{matrix}\right).
\end{equation}
The Wigner 3-$j$ symbol guarantees that two sides of the quadrilateral and the diagonal form a triangle. Given the definition in Eq.~\eqref{eq:connectTrispec}, we denote a particular ordering of the multipoles as
 \begin{align} \label{eq:trispectrumDef}
   \tau^{\ell_1\ell_2\ell_3\ell_4\boldsymbol{X}}_{m_1m_2 m_3 m_4} = \sum_{LM} (-1)^M \begin{pmatrix} \ell_1 &  \ell_2 & L \\ m_1 & m_2 & -M
    \end{pmatrix}  \begin{pmatrix} \ell_3 &  \ell_4 & L \\ m_3 & m_4 & M
    \end{pmatrix} t^{\ell_1\ell_2,\boldsymbol{X}}_{\ell_3 \ell_4}(L),
\end{align}
and observe that the remaining combinations are given by
\begin{align} \label{eq:trispectrum_perms_Def}
   T^{\ell_1\ell_2\ell_3\ell_4,\boldsymbol{X}}_{m_1m_2m_3m_4} =& 
    \tau^{\ell_1\ell_2\ell_3\ell_4,\boldsymbol{X}}_{m_1m_2m_3m_4}+
    \tau^{\ell_2\ell_1\ell_3\ell_4,\boldsymbol{X}}_{m_2m_1m_3m_4}+
    \tau^{\ell_1\ell_2\ell_4\ell_3,\boldsymbol{X}}_{m_1m_2m_4m_3}+
    \tau^{\ell_2\ell_1\ell_4\ell_3,\boldsymbol{X}}_{m_2m_1m_4m_3}+
    \nonumber \\  
    &\tau^{\ell_1\ell_3\ell_2\ell_4,\boldsymbol{X}}_{m_1m_3m_2m_4}+
    \tau^{\ell_1\ell_3\ell_4\ell_2,\boldsymbol{X}}_{m_1m_3m_4m_2}+
    \tau^{\ell_3\ell_1\ell_2\ell_4,\boldsymbol{X}}_{m_3m_1m_2m_4}+
    \tau^{\ell_3\ell_1\ell_4\ell_2,\boldsymbol{X}}_{m_3m_1m_4m_2}+
    \nonumber \\  
    &\tau^{\ell_1\ell_4\ell_3\ell_2,\boldsymbol{X}}_{m_1m_4m_3m_2}+
    \tau^{\ell_4\ell_1\ell_3\ell_2,\boldsymbol{X}}_{m_4m_1m_3m_2} +
    \tau^{\ell_1\ell_4\ell_2\ell_3,\boldsymbol{X}}_{m_1m_4m_2m_3}+
    \tau^{\ell_4\ell_1\ell_2\ell_3,\boldsymbol{X}}_{m_4m_1m_2m_3} .
\end{align}

We may now write the reduced trispectrum as a combination of the radial functions defined in Eq.~\eqref{eq:radial_funcs}. Let us consider the $\tau_\mathrm{NL}$ type of trispectrum of Eq.~\eqref{eq:local_trispectrum_A}. Then 
\begin{equation}
    \begin{split}
     t^{\tau_\mathrm{NL},\boldsymbol{X}}_{\ell_1 \ell_2\ell_3 \ell_4}(L) = \tau_\mathrm{NL}h_{\ell_1 \ell_2 L} h_{\ell_3 \ell_4 L}\int \mathrm{d}r_1 \mathrm{d}r_2  r_1^2r_2^2 \, [&b_L(r_1,r_2)\beta^{X_1}_{\ell_1}(r_1)\alpha^{X_2}_{\ell_2}(r_1)\beta^{X_3}_{\ell_3}(r_2)\alpha^{X_4}_{\ell_4}(r_2)+\\
     &+\text{11 perms.}],
\end{split}
\end{equation}
where we defined 
\begin{equation}
\label{eq:b_L}
    b_L(r_1,r_2) = \frac{2}{\pi}\int \mathrm{d}K\,K^2 j_L(K r_1) j_L(K r_2)P_\zeta (K).
\end{equation}
While Eq.~\eqref{eq:b_L} has no exact solution for a general power spectrum, for a scale invariant power spectrum $P_{\zeta}(K) = 2\pi^2 A_\mathrm{s} K^{-3}$, we find 
\be
 b_L(r_1,r_2) &=& \frac{\pi^2 A_\mathrm{s}}{2}
 \left(
\begin{array}{ccc}
 \frac{2 \left(\frac{r_2}{r_1}\right)^L \Gamma (L) \, _2\tilde{F}_1\left(-\frac{1}{2},L;L+\frac{3}{2};\frac{r_2^2}{r_1^2}\right)}{\sqrt{\pi }} & r_2<r_1 \\
 \frac{2 \left(\frac{r_1}{r_2}\right)^L \Gamma (L) \, _2\tilde{F}_1\left(-\frac{1}{2},L;L+\frac{3}{2};\frac{r_1^2}{r_2^2}\right)}{\sqrt{\pi }} & r_2>r_1 \\
 \frac{4}{\pi} \frac{1}{L(1+L)} & r_2 = r_1\\
\end{array}
\right).
\ee
Here $\tilde{F}_1$ is a hypergeometric function. The solution above is numerically unstable, because for large arguments its value relies on cancellation of large numbers. However, further investigation shows that for $L \leq 10$, $b_L(r_1,r_2)$ is approximately constant over the last scattering surface.
Since this would considerably reduce the computational cost of  $t^{\tau_\mathrm{NL},\boldsymbol{X}}_{\ell_1 \ell_2\ell_3 \ell_4}(L)$, we take this approximation when computing the $S/N$ for this trispectrum. Therefore, the $\tau_\mathrm{NL}$ angular reduced trispectrum is given by
\begin{equation}
    \label{eq:local_ang_trispectrumA}
    \begin{split}
     t^{\tau_\mathrm{NL},\boldsymbol{X}}_{\ell_1 \ell_2\ell_3 \ell_4}(L) = \frac{2\pi A_\mathrm{s}\tau_\mathrm{NL} }{L(1+L)}h_{\ell_1 \ell_2 L} h_{\ell_3 \ell_4 L}\int \mathrm{d}r_1 \mathrm{d}r_2  r_1^2r_2^2 \, [&\beta^{X_1}_{\ell_1}(r_1)\alpha^{X_2}_{\ell_2}(r_1)\beta^{X_3}_{\ell_3}(r_2)\alpha^{X_4}_{\ell_4}(r_2) +\\
     &\text{11 perms.}].
\end{split}
\end{equation}
Along these lines, the $g^\local_\mathrm{NL}$ reduced trispectrum is given by
\begin{equation}
    \begin{split}
     t^{g_\mathrm{NL},\boldsymbol{X}}_{\ell_1 \ell_2\ell_3 \ell_4}(L) =\frac{54}{25} g^\local_\mathrm{NL}h_{\ell_1 \ell_2 L} h_{\ell_3 \ell_4 L} \int& \mathrm{d}r_1 \mathrm{d}r_2  r_1^2r_2^2 \,[a_L(r_1,r_2)\alpha^{X_1}_{\ell_1}(r_1)\beta^{X_2}_{\ell_2}(r_1)\beta^{X_3}_{\ell_3}(r_2)\beta^{X_4}_{\ell_4}(r_2)+\\
     &a_L(r_1,r_2)\beta^{X_1}_{\ell_1}(r_1)\beta^{X_2}_{\ell_2}(r_1)\beta^{X_3}_{\ell_3}(r_2)\alpha^{X_4}_{\ell_4}(r_2)+\text{2 perms.}],
\end{split}
\end{equation}
with
\begin{equation}
\label{eq:a_L}
    a_L(r_1,r_2) = \frac{2}{\pi}\int \mathrm{d}K\,K^2 j_L(K r_1) j_L(K r_2) = \frac{1}{r_1^2}\delta (r_1-r_2),
\end{equation}
Replacing Eq.~\eqref{eq:a_L} in $t^{g_\mathrm{NL},\boldsymbol{X}}_{\ell_1 \ell_2\ell_3 \ell_4}(L)$, then we are left with
\begin{equation}
    \label{eq:local_ang_trispectrumB}
    \begin{split}
     t^{g_\mathrm{NL},\boldsymbol{X}}_{\ell_1 \ell_2\ell_3 \ell_4}(L) =\frac{54}{25} g^\local_\mathrm{NL}h_{\ell_1 \ell_2 L} h_{\ell_3 \ell_4 L} \int& \mathrm{d}r_1r_1^2 \,[\alpha^{X_1}_{\ell_1}(r_1)\beta^{X_2}_{\ell_2}(r_1)\beta^{X_3}_{\ell_3}(r_2)\beta^{X_4}_{\ell_4}(r_2)+\\
     &\beta^{X_1}_{\ell_1}(r_1)\beta^{X_2}_{\ell_2}(r_1)\beta^{X_3}_{\ell_3}(r_2)\alpha^{X_4}_{\ell_4}(r_2)+\text{2 perms.}
     ].
\end{split}
\end{equation}

\section{Cosmic microwave background: flat-sky}
\label{app:cmb_flatsky}
In the flat-sky approximation we ignore the curvature of the sky, which is equivalent to approximate the sphere in the neighborhood of a point by the tangent plane at that point. As a consequence, the spherical harmonic expansion of a perturbation is reduced to a simple Fourier transform~\cite{zaldarriaga:small_limit_cmb,hu:flat_sky}. In this way, it will be easier to appreciate the effect of diffusion damping on the CMB statistics. For simplicity, we consider only temperature anisotropies $\Delta T/T(\hat{\boldsymbol{n}})$. 

In the line-of-sight approach~\cite{seljak:line_of_sight}, these are given by
\begin{equation}
\label{eq:deltaT_lineofsight}
\begin{split}
    \frac{\Delta T}{T}(\hat{\boldsymbol{n}}) &= \int \frac{\mathrm{\mathrm{d}^3}k}{(2\pi)^3}\,\zeta (\bsk) \int_0^{\tau_0} \mathrm{d}\tau \mathrm{e}^{i\bsk\cdot \hat{\boldsymbol{n}}(\tau - \tau_0)}  S(k,\tau),
\end{split}
\end{equation}
where $S(k,\tau)$ is the CMB source function, which encodes all the information about metric perturbations and photon fluctuations, and $\tau$ is the conformal time and $\tau_0$ refers to the conformal time today. The projection on a plane perpendicular to the line of sight is given by~\cite{loverde:limber_approx_flat_sky} 
\begin{equation}
\label{eq:deltaT_perp}
\begin{split}
    \frac{\Delta T}{T}(\hat{\boldsymbol{n}}^\perp) &=  \int \mathrm{d}r\, F(r) \frac{\Delta T}{T}(\hat{\boldsymbol{n}}^\perp,r)\\
    &= \int \mathrm{d}r\,F(r)\int \frac{\mathrm{\mathrm{d}^3}k}{(2\pi)^3}\,\zeta (\bsk) \int_0^{\tau_0} \mathrm{d}\tau \mathrm{e}^{i\bsk\cdot \hat{\boldsymbol{n}}(\tau - \tau_0)}  S(k,\tau),
\end{split}
\end{equation}
with $F(r)$ the projection kernel. Here we consider $F(r)=\delta(r-r_\mathrm{rec})$, where $r_\mathrm{rec}$ is the distance from the last scattering surface. Accordingly, we define the flat-sky temperature anisotropies as
\begin{equation}
\begin{split}
    a(\bl) &= \int \mathrm{\mathrm{d}^2}x^\perp \frac{\Delta T}{T}(\hat{\boldsymbol{n}}^\perp)\mathrm{e}^{-i\bl\cdot \hat{\boldsymbol{n}}^\perp}\\
    &=\int \mathrm{\mathrm{d}^2}\hat{n}^\perp\int \frac{\mathrm{\mathrm{d}^3}k}{(2\pi)^3}\,\zeta (\boldsymbol{k})  \int_0^{\tau_0} \mathrm{d}\tau\,\mathrm{e}^{i[\bsk^\perp(\tau - \tau_0)-\bl]\cdot \hat{\boldsymbol{n}}^\perp}\mathrm{e}^{i k^{\parallel}(\tau - \tau_0)}   S(k,\tau),
\end{split}
\end{equation}
where we used the decomposition $\bsk = (\bsk^\perp, k^\parallel)$ such that $k = \sqrt{(k^\parallel)^2 + \bsk^\perp \cdot \bsk^\perp}$, and $\bl$ is a $2$D wavevector. Finally, since
\begin{equation}
    \int \mathrm{\mathrm{d}^2}\hat{n}^\perp \,\mathrm{e}^{i(\bsk^\perp(\tau_1 - \tau_0)-\bl)\cdot \hat{\boldsymbol{n}}^\perp} = (2\pi)^2 \delta^{(2)} (\bsk^\perp(\tau - \tau_0)-\bl),
\end{equation}
the temperature anisotropies can be rewritten as
\begin{equation}
\label{eq:al_2d_zeta}
\begin{split}
    a(\bl) =\int \frac{\mathrm{\mathrm{d}^3}k}{(2\pi)^3}\,\zeta (\boldsymbol{k})\Delta_\ell^T(k^\parallel),
\end{split}
\end{equation}
with the flat-sky transfer function $\Delta_\ell^T(k^\parallel)$ given by
\begin{equation}
    \Delta_\ell^T(k^\parallel) = (2\pi)^2\int_0^{\tau_0} \mathrm{d}\tau\,\delta^{(2)} (\bsk^\perp(\tau - \tau_0)-\bl)\mathrm{e}^{i k^\parallel (\tau - \tau_0)}   S(k,\tau).
\end{equation}
At this point, we can introduce the flat-sky power spectrum
\begin{equation}
\label{eq:al_2point_correlation}
    \langle a(\bl_1)a(\bl_2)\rangle = (2\pi)^2\delta^{(2)} (\bl_1 + \bl_2) C(\ell_1).
\end{equation}

In the simplest scenario, where the radiation transfer effect is negligible, the temperature anisotropies read as
\begin{equation}
\label{eq:alm_no_damping}
\begin{split}
    a(\bl) = (2\pi)^2\int \frac{\mathrm{\mathrm{d}^3}k}{(2\pi)^3}\,\zeta (\boldsymbol{k})\delta^{(2)} (\bsk^\perp r_\mathrm{rec}-\bl)\mathrm{e}^{i k^\parallel r_\mathrm{rec}}.
\end{split}
\end{equation}
With a few calculations, we obtain $C(\ell) \propto \ell^{-2}$. Given that the scope of this work focuses on the scaling relations, we omitted all the coefficients. In a similar fashion, we derive the flat-sky bispectrum 
\begin{equation}
\label{eq:al_3point_correlation}
    \langle a(\bl_1)a(\bl_2)a(\bl_3)\rangle = (2\pi)^2\delta^{(2)} (\bl_1 + \bl_2+\bl_3) B(\ell_1,\ell_2,\ell_3),
\end{equation}
where
\begin{equation}
    B(\ell_1,\ell_2,\ell_3)= \prod_{i=1}^3\left(\int \frac{\mathrm{\mathrm{d}^3}k_i}{(2\pi)^3}\,\Delta^T_{\ell_i}(k^\parallel_i)\right)\, \delta^{(3)}(\bsk_{123}) B_\zeta(k_1,k_2,k_3).
\end{equation}
As an example, we consider here the local bispectrum of Eq.~\eqref{eq:local_bispectrum}. Replacing Eq.~\eqref{eq:alm_no_damping}, we obtain 
\begin{equation}
\label{eq:undamped_cmb_local_bispectrum}
\begin{split}
      B(\ell_1,\ell_2,\ell_3) \approx A_\mathrm{s}^2f_\mathrm{NL}\left[ \frac{1}{\ell_1^2\ell_2^2}+\text{2 perms.}\right].
\end{split}
\end{equation}
Generally, we may define an angular flat-sky $(N-1)$-spectrum as
\begin{equation}
\label{eq:al_Npoint_correlation}
    \langle a(\bl_1)\cdots a(\bl_N)\rangle = (2\pi)^2\delta^{(2)} (\bl_{1\dots N}) F(\ell_1,\dots,\ell_N).
\end{equation}
The scale invariance of the $(N-1)$-spectrum implies that $F \sim \ell^{-2(N-1)}$. 
\section{The effect of damping on CMB statistics}
\label{app:damping_cmb}
Let us now include the effect of diffusion damping at small scales by introducing~\cite{babich:bispectrum_cmb_polarization,bartolo:ng_from_recombination, bartolo:png_recombination_squeezedlimit}
\begin{equation}
    \label{eq:damping_transfer_function}
    \Delta_\ell^T(k^\parallel) \approx  \delta^{(2)} (\bsk^\perp(\tau_\mathrm{rec}  - \tau_0)-\bl)
    \mathrm{e}^{-1/2(\ell/\ld)^{1.2}}\mathrm{e}^{-1/2(k^\parallel/k_\mathrm{D})^{1.2}},
\end{equation}
where $\ld\simeq 1300$ is the damping scale, namely the scale above which diffusion damping dominates over gravitational pull, and the exponential cuts the integral off at the corresponding damping mode $k_\mathrm{D}$. The exponent $1.2$ comes from the study of Ref.~\cite{hu:dampingtail_cmb} and is approximate. Its precise value is irrelevant for the work presented in this paper. This transfer function highlights the fact that the damping effect is 3-dimensional: it has a $k^\perp \propto \ell$ and a line of sight $k^\parallel$ component. The former integrates out, effectively becoming a multiplicative transfer function term. However, the line of sight damping, which probes the thickness of the last scattering surface, remains a convolution, coupling $k^\parallel$ modes. The net result is a change in the scaling behavior of the correlators.

Indeed, using Eq.~\eqref{eq:damping_transfer_function} in the power spectrum expression, we obtain
\begin{equation}
    \begin{split}
        C(\ell) \approx \frac{A_\mathrm{s}}{\ell^2}\frac{\mathrm{e}^{-(\ell/\ld)^{1.2}}}{\sqrt{1+(\ell/\ld)^2}} \approx \frac{A_\mathrm{s}}{\ell^3}\mathrm{e}^{-(\ell/\ld)^{1.2}}
    \end{split}
\end{equation}
where the last passage holds for $\ell\gg \ld$.  As expected, the angular power spectrum changes scaling, from $\ell^{-2}$ to $\ell^{-3}$. 
Likewise, for the local bispectrum it is straightforward to show that 
\begin{equation}
\begin{split}
    B(\ell_1,\ell_2,\ell_3) \approx A_\mathrm{s}^2f_\mathrm{NL}\mathrm{e}^{-(\ell_1^{1.2}+\ell_2^{1.2}+\ell_3^{1.2})/\ld^{1.2}}\left[ \frac{1}{\ell_1^2\sqrt{1+(\ell_1/\ld)^2}}\frac{1}{\ell_2^2\sqrt{1+(\ell_2/\ld)^2}}+\text{2 perms.}\right]
\end{split}
\end{equation}
and for $\ell\gg \ld$ we obtain
\begin{equation}
\begin{split}
    B(\ell_1,\ell_2,\ell_3) \approx A_\mathrm{s}^2f_\mathrm{NL}\mathrm{e}^{-(\ell_1^{1.2}+\ell_2^{1.2}+\ell_3^{1.2})/\ld^{1.2}}\left[ \frac{1}{\ell_1^3\ell_2^3}+\text{2 perms.}\right]
\end{split}
\end{equation}
In general, the expected scaling can be obtained by a simple replacement $\ell^2\rightarrow \ell^2(\ell/\ld)$ and by including the exponential factor $\mathrm{e}^{-\sum_i(\ell_i/\ld)^{1.2}}$. However, the exponential factor cancels out in the $S/N$ computation, since it appears equally in the numerator and the denominator.

\section{Optimal estimator for non-Gaussianity and signal-to-noise} 
\label{app:snr_derivation}
\subsection{Large scale structure}
\label{app:snr_estimation_3d}
Optimal estimators for the amplitude of non-Gaussianity can be constructed from the bispectrum and trispectrum of the primordial perturbations, and extended to the general $(N-1)$-spectrum~\eqref{eq:moments_curvature}. 

In the weak non-Gaussian limit and assuming statistical isotropy and homogeneity, it has been shown that the optimal estimator coming from the bispectrum is given by~\cite{sefusatti:bispectrum_nbody_sim,fergusson:separable_bispectrum_lss}
\begin{equation}
    \begin{split}
        \widehat{\mathcal{E}}^{(3)} = \frac{1}{\mathcal{N}_{(3)}}\int \frac{\mathrm{d}^3k_1}{(2\pi)^3}\frac{\mathrm{d}^3k_2}{(2\pi)^3}\frac{\mathrm{d}^3k_3}{(2\pi)^3}\,\frac{\langle\zeta^\mathrm{th}_{\mathbf{k}_1}\zeta^\mathrm{th}_{\mathbf{k}_2}\zeta^\mathrm{th}_{\mathbf{k}_3}\rangle}{P(k_1)P(k_2)P(k_3)}[\zeta_{\mathbf{k}_1}\zeta_{\mathbf{k}_2}\zeta_{\mathbf{k}_3}-3\langle\zeta_{\mathbf{k}_1}\zeta_{\mathbf{k}_2}\rangle\zeta_{\mathbf{k}_3}].
    \end{split}
    \label{eq:bispectrum_estimator}
\end{equation}
Here the primordial curvature perturbation $\zeta_{\mathbf{k}}$ comes from observations, while $\zeta^\mathrm{th}_{\mathbf{k}}$ is the theoretically motivated one. The linear term $\langle\zeta_{\mathbf{k}_1}\zeta_{\mathbf{k}_2}\rangle\zeta_{\mathbf{k}_3}$ accounts for any anisotropic effect due to systematics in the dataset. This terms is irrelevant for the simple Fisher forecasts we present here, and we will neglect this term from hereon. 
For the estimator to be unbiased, the normalization factor $\mathcal{N}$ must be given by the inverse Fisher information
\begin{equation}
\begin{split}
    \mathcal{N} = F^{-1} \equiv V (2\pi)^3\int \frac{\mathrm{d}^3k_1}{(2\pi)^3}\frac{\mathrm{d}^3k_2}{(2\pi)^3}\frac{\mathrm{d}^3k_3}{(2\pi)^3}\,\delta^{(3)}(\bsk_{123})\frac{B^2_\zeta(k_1,k_2,k_3)}{P_\zeta(k_1)P_\zeta(k_2)P_\zeta(k_3)},
\end{split}
\end{equation}
where $P(k)$ and $B_\zeta(k_1,k_2,k_3)$ are the power spectrum and the bispectrum respectively (Eq.~\eqref{eq:moments_curvature}), and $V$ is the volume of the survey. 

Since an optimal estimator also saturates the Cram\`er-Rao inequality, its minimum error is given by the Fisher information itself, thus the signal-to-noise ratio
\begin{equation}
\label{eq:estimator_variance}
    \sigma(\widehat{\mathcal{E}}) = F^{-1/2} = (S/N)^{-1}.
\end{equation}
From now on we will use only the $S/N$, 
\begin{equation}
      \begin{split}
       \left(\frac{S}{N}\right)^2_{(3)}= V (2\pi)^3\int \frac{\mathrm{d}^3k_1}{(2\pi)^3}\frac{\mathrm{d}^3k_2}{(2\pi)^3}\frac{\mathrm{d}^3k_3}{(2\pi)^3}\,\delta^{(3)}(\bsk_{123})\frac{B^2_\zeta(k_1,k_2,k_3)}{P(k_1)P(k_2)P(k_3)},
    \end{split}
    \label{eq:snr_lss_diracdelta}
\end{equation}
with its connection to the variance of the estimator explicit in Eq.~\eqref{eq:estimator_variance}.

Likewise, we define the trispectrum optimal estimator
\begin{equation}
    \begin{split}
        \widehat{\mathcal{E}}^{(4)} = \frac{1}{\mathcal{N}_{(4)}}\int \frac{\mathrm{d}^3k_1}{(2\pi)^3}\frac{\mathrm{d}^3k_2}{(2\pi)^3}\frac{\mathrm{d}^3k_3}{(2\pi)^3}\frac{\mathrm{d}^3k_4}{(2\pi)^3}\,&\frac{\langle\zeta^\mathrm{th}_{\mathbf{k}_1}\zeta^\mathrm{th}_{\mathbf{k}_2}\zeta^\mathrm{th}_{\mathbf{k}_3}\zeta^\mathrm{th}_{\mathbf{k}_4}\rangle_c}{P(k_1)P(k_2)P(k_3)P(k_4)}[\zeta_{\mathbf{k}_1}\zeta_{\mathbf{k}_2}\zeta_{\mathbf{k}_3}\zeta_{\mathbf{k}_4})-\\
        &-6\langle\zeta_{\mathbf{k}_1}\zeta_{\mathbf{k}_2}\rangle\zeta_{\mathbf{k}_3}\zeta_{\mathbf{k}_4}+3\langle\zeta_{\mathbf{k}_1}\zeta_{\mathbf{k}_2}\rangle\langle\zeta_{\mathbf{k}_3}\zeta_{\mathbf{k}_4})],
    \end{split}
\end{equation}
and the $(N-1)$-spectrum estimator
\begin{equation}
    \begin{split}
        \widehat{\mathcal{E}}^{(N)} = \frac{1}{\mathcal{N}_{(N)}}\int \frac{\mathrm{d}^3k_1}{(2\pi)^3}&\frac{\mathrm{d}^3k_2}{(2\pi)^3}\cdots\frac{\mathrm{d}^3k_N}{(2\pi)^3}\,\frac{\langle\zeta^\mathrm{th}_{\mathbf{k}_1}\zeta^\mathrm{th}_{\mathbf{k}_2}\cdots\zeta^\mathrm{th}_{\mathbf{k}_N}\rangle_c}{P(k_1)P(k_2)\cdots P(k_N)}[\zeta_{\mathbf{k}_1}\zeta_{\mathbf{k}_2}\cdots \zeta_{\mathbf{k}_N}-\\
        &-A\langle\zeta_{\mathbf{k}_1}\zeta_{\mathbf{k}_2}\rangle\zeta_{\mathbf{k}_3}\cdots\zeta_{\mathbf{k}_N}+B\langle\zeta_{\mathbf{k}_1}\zeta_{\mathbf{k}_2}\rangle\langle\zeta_{\mathbf{k}_3}\zeta_{\mathbf{k}_4}\rangle\zeta_{\mathbf{k}_5}\cdots\zeta_{\mathbf{k}_N}+\dots],
    \end{split}
\end{equation}
where $A$ and $B$ are coefficient that take into account permutations of terms, and the dots refer to products of two-point correlation functions that arise as $N$ increases. These terms are necessary to generalize to the case of incomplete sample coverage, inhomogeneous noise and removing any Gaussian noise bias . However for the purpose of this a simple Fisher forecast, we can neglect them.

Using Eq.~\eqref{eq:estimator_variance}, the $(S/N)^2$ of the trispectrum is given by
\begin{equation}
      \begin{split}
       \left(\frac{S}{N}\right)^2_{(4)}= V^{(4)} (2\pi)^3\int \frac{\mathrm{d}^3k_1}{(2\pi)^3}\frac{\mathrm{d}^3k_2}{(2\pi)^3}\frac{\mathrm{d}^3k_3}{(2\pi)^3}\frac{\mathrm{d}^3k_4}{(2\pi)^3}\,\delta^{(3)}(\bsk_{1234})\frac{T^2_\zeta(k_1,k_2,k_3,k_4)}{P(k_1)P(k_2)P(k_3)P(k_4)},
    \end{split}
    \label{eq:snr_lss_trispectrum}
\end{equation}
and in the case of $(N-1)$-spectrum we obtain
\begin{equation}
      \begin{split}
       \left(\frac{S}{N}\right)^2_{(N)}= V^{(N)} (2\pi)^3\int \frac{\mathrm{d}^3k_1}{(2\pi)^3}\frac{\mathrm{d}^3k_2}{(2\pi)^3}\cdots\frac{\mathrm{d}^3k_N}{(2\pi)^3}\,\delta^{(3)}(\bsk_{12\dots N})\frac{F^2_\zeta(k_1,k_2,\dots,k_N)}{P(k_1)P(k_2)\cdots P(k_N)},
    \end{split}
    \label{eq:snr_lss_n_spectrum}
\end{equation}

\subsection{Cosmic microwave background: full-sky analysis}
\label{app:cmb_snr_derivation_fullsky}
Similarly to the LSS case, we can define an optimal estimator to extract non-Gaussian information from CMB data. Let us consider the bispectrum optimal estimator, given by~\cite{creminelli:ng_from_wmap}
\begin{equation}
    \label{eq:fnl_optimal_estimator}
    \begin{split}
        \widehat{\mathcal{E}}^{(3)}  = \frac{1}{F^{-1}} &\sum_{X_i, X'_i} \sum_{\ell_i,m_i}
 \sum_{\ell'_i, m'_i}\mathcal{G}^{\ell_1\ell_2\ell_3}_{m_1 m_2 m_3 }
 b^{X_1 X_2 X_3, \mathrm{th}}_{\ell_1 \ell_2 \ell_3}\times\\
  \times&\bigg\{\left[\left(\mathbb{C}^{-1}_{\ell_1 m_1, \ell_1' m_1'}\right)^{X_1 X'_1}
 a^{X'_1}_{\ell_1'm_1'} \left(\mathbb{C}^{-1}_{\ell_2 m_2, \ell_2' m_2'} \right)^{X_2 X'_2}
 a^{X'_2}_{\ell_2'm_2'}\left(\mathbb{C}^{-1}_{\ell_3 m_3, \ell_3' m_3'}\right)^{X_3 X'_3}
 a^{X'_3}_{\ell_3'm_3'} \right]-\\
 &-\left[ \left(\mathbb{C}^{-1}_{\ell_1 m_1, \ell_2 m_2}\right)^{X_1 X_2}
 \left(\mathbb{C}^{-1}_{\ell_3 m_3,\ell_3' m_3'}\right)^{X_3 X'_3} a^{X'_3}_{\ell_3'm_3'}
  + \mathrm{cyclic} \right] \bigg\}, 
\end{split}
\end{equation}
where $ b_{\ell_1 \ell_2 \ell_3}^{\rm th}$ is the theoretical reduced bispectrum and $a^{X_1}_{\ell_1 m_1}$ are observed/simulated multipoles and $F^{-1}$ the inverse Fisher information. $\mathbb{C}^{-1}$ is the inverse of the covariance matrix, which is given by a block matrix
\begin{equation}
\mathbb{C}=\left(
\begin{array}{cc}
\mathbb{C}^{TT} & \mathbb{C}^{TE} \\

\mathbb{C}^{ET} & \mathbb{C}^{EE}
\end{array}\right),
\end{equation}
where the blocks represent the full $TT$, $TE$, and $EE$ covariance matrices, with $\mathbb{C}^{ET}$ being the transpose of $\mathbb{C}^{TE}$. We neglect the linear term in the estimator~\eqref{eq:fnl_optimal_estimator} since it is proportional to the monopole when rotational invariance is valid. We also work in the ``diagonal covariance'' approximation~\cite{yadav:fast_estimator_png_TE}, which reduces the computation of the inverse covariance matrix to a $2\times 2$ matrix
\begin{equation}
\label{eq:cmb_covariance}
   \mathbb{C}_\ell = \begin{pmatrix}
    C^{TT}_\ell && C^{TE}_\ell\\
    C^{ET}_\ell && C^{EE}_\ell
    \end{pmatrix},
\end{equation}
and $C_\ell$ is the CMB power spectrum. Thus, the estimator~\eqref{eq:fnl_optimal_estimator} reads as
\begin{equation}
    \label{eq:fnl_optimal_estimator_simple}
    \begin{split}
        \widehat{\mathcal{E}}^{(3)}  =  & \frac{1}{F^{-1}} \sum_{X_i, X'_i} \sum_{\ell_i,m_i} \mathcal{G}^{\ell_1\ell_2\ell_3}_{m_1 m_2 m_3 }b^{X_1 X_2 X_3, \mathrm{th}}_{\ell_1 \ell_2 \ell_3} (\mathbb{C}^{-1})_{\ell_1}^{X_1 X'_1} (\mathbb{C}^{-1})_{\ell_2} ^{X_2 X'_2}  (\mathbb{C}^{-1})_{\ell_3}^{X_3 X'_3} a^{X'_1}_{\ell_1 m_1}a^{X'_2}_{\ell_2 m_2}a^{X'_3}_{\ell_3 m_3}.
\end{split}
\end{equation}

Since the estimator~\eqref{eq:fnl_optimal_estimator} is optimal, the variance on it is given by $\sigma( \widehat{\mathcal{E}}^{(3)}) = (F^{-1})^{1/2} = (S/N)^{-1}$, with the $(S/N)^2$ given by
\begin{equation}
    \left(\frac{S}{N}\right)^2_{(3)} = \sum_{X_i, X'_i} \sum_{\ell_i} h^{\ell_1\ell_2\ell_3}b^{X_1 X_2 X_3}_{\ell_1 \ell_2 \ell_3} (\mathbb{C}^{-1})_{\ell_1}^{X_1 X'_1} (\mathbb{C}^{-1})_{\ell_2} ^{X_2 X'_2}  (\mathbb{C}^{-1})_{\ell_3}^{X_3 X'_3}b^{X'_1 X'_2 X'_3}_{\ell_1 \ell_2 \ell_3},
\end{equation}
where
 \begin{equation}
    h^{\ell_1\ell_2\ell_3} = \sum_{m_i} (\mathcal{G}^{\ell_1\ell_2\ell_3}_{m_1 m_2 m_3 })^2 = \frac{(2\ell_1+1)(2\ell_2+1)(2\ell_3+1)}{4\pi}\begin{pmatrix} \ell_1 & \ell_2 & \ell_3 \\ 0 & 0 & 0 \end{pmatrix}^2. 
\end{equation}
Here, the Wigner-$3j$ matrix in $h^{\ell_1\ell_2\ell_3}$ enforces the triangle condition between the triplet $(\ell_1,\ell_2,\ell_3)$. In that sense, the domain of calculation is a tetrahedron. However, since the problem is symmetric under permutation of indices, we can assume $\ell_1\leq\ell_2\leq\ell_3$ without any loss in generality and with a significant gain in computational speed. Therefore, the $(S/N)^2$ becomes
\begin{equation}
\label{eq:cmb_fnl_snr}
    \left(\frac{S}{N}\right)^2_{(3)} = \sum_{X_i, X'_i} \sum_{\ell_1\leq\ell_2\leq\ell_3} \frac{h^{\ell_1\ell_2\ell_3}}{f(\ell_1,\ell_2,\ell_3)} b^{X_1 X_2 X_3}_{\ell_1 \ell_2 \ell_3} (C_{\ell_1}^{X_1 X'_1})^{-1} (C_{\ell_2}^{X_2 X'_2})^{-1} (C_{\ell_3}^{X_3 X'_3})^{-1}b^{X'_1 X'_2 X'_3}_{\ell_1 \ell_2 \ell_3}.
\end{equation}
The function $f(\ell_1,\ell_2,\ell_3)$ takes values 1, 2, and 6 when all $\ell$'s are different, two of them are same, and all are same, respectively. 

Likewise, we define an optimal estimator for the trispectrum as~\cite{regan:trispectrum_estimation}
\be
\label{eq:trispectrum_optimal_estimator}
\widehat{\mathcal{E}}^{(4)}  &=& \frac{1}{F^{-1}} \sum_{X_i, X'_i}\sum_{\ell_i,m_i}
 \langle a_{\ell_1 m_1}^{X_1}a_{\ell_2 m_2}^{X_2} a_{\ell_3 m_3}^{X_3} a_{\ell_4 m_4}^{X_4}	\rangle_c\bigg[\left(\mathbb{C}^{-1} \right)_{\ell_1}^{X_1 X'_1}
 \left(\mathbb{C}^{-1} \right)_{\ell_2}^{X_2 X'_2}\left(\mathbb{C}^{-1} \right)_{\ell_3}^{X_3 X'_3} \left(\mathbb{C}^{-1} \right)_{\ell_4}^{X_4 X'_4}\times \nonumber\\
 &&\times 
 a_{\ell_1m_1}^{X'_1}a_{\ell_2m_2}^{X'_2}a_{\ell_3m_3}^{X'_3}a_{\ell_4m_4}^{X'_4}\bigg],
\ee
where again we drop the linear terms. 

As for the bispectrum estimator, in order to have an unbiased estimator, the normalization must be the inverse Fisher matrix. Then the estimator variance, the $(S/N)^2$, reads
\be
\label{eq:cmb_trispectrum_snr}
\left(\frac{S}{N}\right)^2_{(4)} &=& \sum_{X_i, X'_i}\sum \limits_{\ell_i} \sum \limits_L \frac{1}{(2L+1)} T^{X_1 X_2 X_3 X_4}_{c,\ell_1\ell_2\ell_3\ell_4}(L)\left(\mathbb{C}^{-1} \right)_{\ell_1}^{X_1 X'_1}\left(\mathbb{C}^{-1} \right)_{\ell_2}^{X_2 X'_2}\times  \nonumber \\
&& \left(\mathbb{C}^{-1} \right)_{\ell_3}^{X_3 X'_3} \left(\mathbb{C}^{-1} \right)_{\ell_4}^{X_4 X'_4}T^{X'_1 X'_2 X'_3 X'_4}_{c,\ell_1\ell_2\ell_3\ell_4}(L).
\ee
The $L$ term at the denominator comes from the summation of Wigner-$3j$ matrices
\begin{equation}
    \begin{split}
        &\sum_{m_1m_2} \begin{pmatrix} \ell_1 & \ell_2 & L \\ m_1 & m_2 & -M \end{pmatrix}\begin{pmatrix} \ell_1 & \ell_2 & L' \\ m_1 & m_2 & -M' \end{pmatrix} = \frac{\delta_{LL'}\delta_{MM'}}{2L+1},\\
        &\sum_{M}\sum_{m_3 m_4} \begin{pmatrix} \ell_3 & \ell_4 & L \\ m_3 & m_4 & M \end{pmatrix} \begin{pmatrix} \ell_3 & \ell_4 & L \\ m_3 & m_4 & M \end{pmatrix} = 1.
    \end{split}
\end{equation}

\subsection{Cosmic microwave background: flat-sky analysis}
The $S/N$ in the flat-sky approximation can be easily derived from the full-sky definition~\eqref{eq:cmb_fnl_snr} (for temperature only) by replacing 
\begin{equation}
    \sum_{\ell_i} \rightarrow \int \mathrm{d^2}\ell_i,\quad h^{\ell_1\ell_2\ell_3}b_{\ell_1 \ell_2 \ell_3} \rightarrow \delta^{(2)}(\bl_{123})B(\ell_1, \ell_2, \ell_3),\quad C^{-1}_{\ell_i} \rightarrow C^{-1}(\ell_i).
\end{equation}
Therefore, for the bispectrum we have
\begin{equation}
\begin{split}
    \left(\frac{S}{N}\right)^2_{(3)} = \frac{f_\mathrm{sky}}{6\pi}\frac{1}{(2\pi)^2}\int \mathrm{d}^2\ell_1\mathrm{d}^2\ell_2\mathrm{d}^2\ell_3
    \delta^{(2)}(\bl_{123}) \frac{B^2(\ell_1,\ell_2,\ell_3)}{ C(\ell_1)C(\ell_2)C(\ell_3)},
\end{split}
\label{eq:flatsky_snr_bispectrum}
\end{equation}
where $f_\mathrm{sky}$ refers to the portion of the observed sky. Along these lines we can define the trispectrum flat-sky $S/N$ as 
\begin{equation}
\begin{split}
    \left(\frac{S}{N}\right)^2_{(4)} = \frac{f_\mathrm{sky}}{6\pi}\frac{1}{(2\pi)^2}\int \mathrm{d}^2\ell_1\mathrm{d}^2\ell_2\mathrm{d}^2\ell_3\mathrm{d}^2\ell_4
    \delta^{(2)}(\bl_{1234}) \frac{T^2(\ell_1,\ell_2,\ell_3,\ell_4)}{ C(\ell_1)C(\ell_2)C(\ell_3)C(\ell_4)},
\end{split}
\label{eq:flatsky_snr_trispectrum}
\end{equation}
and the  $S/N$ from the $N$-point correlation function 
\begin{equation}
\begin{split}
    \left(\frac{S}{N}\right)^2_{(N)} = \frac{f_\mathrm{sky}}{6\pi}\frac{1}{(2\pi)^2}\int \mathrm{d}^2\ell_1\mathrm{d}^2\ell_2\cdots\mathrm{d}^2\ell_N
    \delta^{(2)}(\bl_{12\dots N}) \frac{F^2(\ell_1,\ell_2,\dots,\ell_N)}{ C(\ell_1)C(\ell_2)\cdots C(\ell_N)}.
\end{split}
\label{eq:flatsky_snr_n_spectrum}
\end{equation}

\section{Multiple squeezed and collapsed limits}
\label{app:multiple_squeezed_collapsed}
In this appendix, we derive the effect of multiple squeezed or collapsed limits on the $S/N$. In our numerical analysis we encountered the $g_\mathrm{NL}$-trispectrum, where the scaling of the $S/N$ is determined by nested squeezed limits which results in a $S/N$ that scales as mode-counting. We will show that nested squeezed limits never exceed mode counting (besides the $\log$-enhancement), whereas, as already seen in Sec.~\ref{subsec:collapsed_limit} for a single collapsed limit, multiple collapsed limits can go beyond mode counting. 
Our calculations are highly idealistic and simplifying, therefore the enhanced scalings should be carefully interpreted within a specific model.

First, we consider the double squeezed limit of the damped trispectrum\footnote{The higher number of squeezed limits does not affect the scaling of $3$D and undamped $2$D spectra.}
\begin{equation}
    \lim_{\ell_1\ll \ell_2 \ll \ell_{3,4}} T(\ell_1,\ell_2,\ell_3,\ell_4)\propto \frac{\ld^3}{(\ell_1 \ell_2 \ell_3)^{3}} \left(\frac{\ell_1 \ell_2}{\ell_3^2}\right)^{\Delta},
\end{equation}
then the leading scaling in $\lmax$ is given by
\begin{equation}
     \begin{split}
         \left(\frac{S}{N}\right)_{(4)}^2\sim \ld^2 \int_{\ld}^{c \lmax} d^2\ell_{1} d^2\ell_{2} \int_{c \lmax}^{\lmax} d^2\ell_{3}\frac{\ell_1^{-6+2\Delta}\ell_2^{-6+2\Delta}\ell_3^{-6-4\Delta}}{\ell_1^{-3}\ell_2^{-3}\ell_3^{-6}}  
        \sim \lmax^{2-4\Delta},\quad \Delta>0.
         \end{split}
\end{equation}
As shown in Sec.~\ref{subsec:squeezed_limit}, for local shapes, i.e. $\Delta= 0$, the signal-to-noise scales as $\lmax^2$. In general, there is an improvement with respect to the expected scaling $\log \lmax$ for values $0\leq \Delta < 1/2$.

It is possible to extend the argument to $S$ squeezed limits, $\ell_1\ll \ell_2 \ll\dots\ll \ell_S \ll \ell_{S+1}\sim\dots \sim\ell_N$. Notice that the maximum number of squeezed limits is $S_\mathrm{max} = N-2$. In this limit, the $(N-1)$-spectrum reads
\begin{equation}
    F_\zeta(\ell_1,\dots,\ell_N)\sim \frac{\ld^{N-1}}{\ell_1^{3S}\ell^{3(N-S-1)}} \left(\frac{\ell_1}{\ell}\right)^{S\Delta}\sim \ld^{N-1} \ell_1^{-3S+S\Delta}\ell_3^{-3(N-S-1)-S\Delta},
\end{equation}
therefore the $(S/N)^2$ scales as
\begin{equation}
     \begin{split}
         \left(\frac{S}{N}\right)^2\sim \int_{\ld}^{c \lmax} d^{2S}\ell_1 \int_{c \lmax}^{\lmax} d^{2(N-1-S)}\ell\,\frac{\ell_1^{-6S+2S\Delta}\ell^{-6(N-S-1)-2S\Delta}}{\ell_1^{-3S}\ell^{-3(N-S)}} 
        \sim \lmax^{4-N+S(1-2\Delta)}.
         \end{split}
\end{equation}
When $S=1$, we recover the single-squeezed limit scaling $\lmax^{5-N-2\Delta}$. Moreover, we notice that in the interval of values $0 < \Delta \leq 1/2$, the scaling gets enhanced. Indeed, assuming that $(S/N)^2 \propto \lmax^p$, the multiple squeezed limit allows values of $p>0$ for $N<S(1-2\Delta)+4$. Knowing that the possible values of $S$ are between $0\leq S \leq N-2$, we may conclude that the maximum scaling attainable is $p_\mathrm{max} = 2-2\Delta (N-2)$, therefore it never exceeds mode counting.

The computation of multiple collapsed limits follows trivially from the result for a singularly collapsed computation in Eq.~\eqref{eq:snr_collapsed_computation}. In this expression, the integrals from the left and right momenta have the exact form of a product of the $(S/N)^2$ for a $A+1$ spectrum and a $B+1$ spectrum. Thus the scaling for multiply collapsed (or mixed squeezed-collapsed) can be obtained by multiplying the limits for the $A+1$ spectrum by the $B+1$ spectrum. Further, we note that this can be repeated in a nested fashion. E.g., consider a double collapsed $(N-1)$-spectrum in $3$D, then
\begin{align}
    \left(\frac{S}{N}\right)^2_{(N)}\sim &\int_{\kmin}^{c \kmax}\mathrm{d}^3 k_I k_I^{-6+4\Delta}\int_{\kmin}^{c \kmax}\mathrm{d}^3 k_{I'} k_{I'}^{-6+4\Delta} \int_{c\kmax}^{ \kmax}\mathrm{d}^{3A} k_R \frac{k_R^{-6A-4\Delta}}{k_R^{-3(A+1)}} \nonumber \\ & \times\int_{c\kmax}^{ \kmax}\mathrm{d}^{3B_1} k_{R'}\, \frac{k_{R'}^{-6B_1-2\Delta}}{k_{R'}^{-3(B_1+1)}} \times\int_{c\kmax}^{ \kmax}\mathrm{d}^{3B_2} k_{L'}\, \frac{k_{L'}^{-6B_2-2\Delta}}{k_{L'}^{-3(B_2+1)}}\sim \kmax^{9-8\Delta}.
\end{align}
where we split $B = B_1 +B_2$. Here, we expect an enhanced scaling for $\Delta < 3/4$. As for the multiple squeezed limits, we can extend the argument to general $C$ nested collapsed limits where we obtain $(S/N)^2\sim \kmax^{3+C(3-4\Delta)}$. In an analogous manner, for the undamped CMB $(N-1)$-spectrum we find $(S/N)^2\sim \lmax^{2+2C(1-4\Delta/3)}$, while the damped regime produces a scaling given by
\begin{align}
    \left(\frac{S}{N}\right)^2_{(N)}\sim \lmax^{4-N+4C(1-\Delta)}.
\end{align}
We recover the scaling for a single collapsed limit, $(S/N)^2\sim \lmax^{8-N-4\Delta}$. Unlike the case of multiple squeezed limits, the maximum number of nested collapsed limits $C_{\rm max}(N)$ depends on the specific $N$-point correlation function considered, e.g. $C_\mathrm{max}(6) = C_\mathrm{max}(7) = 2$, while the $8$-point correlation function has $C_\mathrm{max}(8) = 3$. 

Note that while these scalings might look impressive (in particular for an undamped tracer), as for the single collapsed limit, they must be critically assessed within a model.
It will be challenging to verify these enhanced scalings in a full Fisher analysis, as the naive computational cost to evaluate $S/N$ of the $N$-point correlator scales as $\ell_{\rm max}^{2N-1}$.

\bibliographystyle{utcaps}
\bibliography{References}

\end{document}